\documentclass[footinbib, aps,prl,reprint, 10pt,twocolumn,showpacs, superscriptaddress]{revtex4-1}
\usepackage{amsfonts,amsmath,amssymb}
\usepackage{blindtext}
\usepackage{graphicx}
\usepackage[pdfstartview=FitH,colorlinks=true,linkcolor=blue,citecolor=blue,urlcolor=blue]{hyperref}
\usepackage{tikz}
\usepackage{verbatim}

\newcommand{\lettersection}[1]{\paragraph*{#1.---}}

\usepackage{times}

\usepackage[english]{babel}
\usepackage{latexsym}
\usepackage{graphics}
\usepackage{subfigure}
\usepackage{epsfig}
\usepackage{color}
\usepackage{braket}
\usepackage[T1]{fontenc}
\usepackage[latin9]{inputenc}
\usepackage{amstext}
\usepackage{esint}
\usepackage[normalem]{ulem}   

\newcommand{\beq}{\begin{equation}}
\newcommand{\eeq}{\end{equation}}

\begin{document}

\title{Thermal fading of the $1/k^4$-tail of the momentum distribution induced by the hole anomaly}

\author{Giulia De Rosi}
\email{giulia.de.rosi@upc.edu}
\affiliation{Departament de F\'isica, Universitat Polit\`ecnica de Catalunya, Campus Nord B4-B5, 08034 Barcelona, Spain}

\author{Grigori E. Astrakharchik}
\email{grigori.astrakharchik@upc.edu}
\affiliation{Departament de F\'isica, Universitat Polit\`ecnica de Catalunya, Campus Nord B4-B5, 08034 Barcelona, Spain}
\affiliation{Departament de F{\'i}sica Qu{\`a}ntica i Astrof{\'i}sica, Facultat de F{\'i}sica, Universitat de Barcelona, E-08028 Barcelona, Spain}

\author{Maxim Olshanii}
\affiliation{Department of Physics, University of Massachusetts Boston, Boston Massachusetts 02125, USA}

\author{Jordi Boronat}
\email{jordi.boronat@upc.edu}
\affiliation{Departament de F\'isica, Universitat Polit\`ecnica de Catalunya, Campus Nord B4-B5, 08034 Barcelona, Spain}

\date{\today}

\begin{abstract}

We study the thermal behavior of correlations in a one-dimensional Bose gas with tunable interaction strength, crossing from weakly-repulsive to Tonks-Girardeau regime.~A reference temperature in this system is that of the hole anomaly, observed as a peak in the specific heat and a maximum in the chemical potential.~We find that at large momenta $k$ and temperature above the anomaly threshold, the tail $\mathcal{C}/k^4$ of the momentum distribution (proportional to the Tan contact $\mathcal{C}$)  is screened by the $1/|k|^3$-term due to a dramatic thermal increase of the internal energy emerging from the thermal occupation of spectral excitation states.~The same fading is consistently revealed in the behavior at short distances $x$ of the one-body density matrix (OBDM) where the $|x|^3$-dependence disappears for temperatures above the anomaly.~We obtain a new general analytic tail for the momentum distribution and a minimum $k$ fixing its validity range, both calculated with exact Bethe-Ansatz method and valid in all interaction and thermal regimes, crossing from the quantum to the classical gas limit.~Our predictions are confirmed by comparison with ab-initio Path Integral Monte Carlo calculations for the momentum distribution and the OBDM exploring a wide range of interaction strength and temperature.~Our results unveil a novel connection between excitations and correlations.~We expect them to be of interest to any cold atomic, nuclear, solid-state, electronic and spin system exhibiting an anomaly or a thermal second-order phase transition. 
\end{abstract}

\maketitle

Tan relation $n\left(k\right) \sim \mathcal{C}/k^4$ describes the tail of the momentum distribution $n\left(k\right)$ at high momenta $k$ and its amplitude is fixed by the Tan contact parameter $\mathcal{C}$~\cite{Tan2008, Tan2008II, Tan2008III}.~This universal law is valid for a broad range of quantum systems, from nucleons~\cite{Bulgac2023} to ultracold atoms~\cite{Olshanii2003, Barth2011}, of bosonic and fermionic statistics, with any interaction strength and particle number~\cite{Tan2008III}.~It applies to multicomponent systems~\cite{Matveeva_2016, Patu2017} and in arbitrary conditions of confinement~\cite{Minguzzi2002, Olshanii2003, Rigol2005} and spatial dimensionality~\cite{Barth2011, Werner2012, Werner2012b}.~It provides a key connection between microscopic large-momenta (short-distance) correlations and macroscopic thermodynamic quantities such as $\mathcal{C}$~\cite{Braaten2012}.

Tan relation is based on the assumption that the tail of the momentum distribution depends entirely on contact two-body interactions \cite{Tan2008, Tan2008II, Tan2008III, Braaten2012} which are modelled only by the universal $s$-wave scattering length \cite{Pitaevskii2016} entering into $\mathcal{C}$.~$n\left(k\right) \sim \mathcal{C}/k^4$ holds then whenever the interaction range $r_0$ is negligible compared to all other relevant length scales of the problem, including the average interparticle distance $d$ and the thermal de Broglie wavelength $\lambda = \sqrt{2 \pi \hbar^2/\left(m k_B T\right)}$.~Tan law is valid for momenta much larger than the average momentum of particles $d^{-1}, \lambda^{-1} \ll |k| \ll r_0^{-1}$.~The present work provides a precise estimate for the minimum momentum $k_{\rm min}$ above which the tail of the distribution is defined $k_{\rm min} \lesssim |k|$.~This new $k_{\rm min}$ holds for any interaction strength and temperature.

Tan relation was believed to be well justified in ultracold atomic gases due to their extreme diluteness and low temperatures.~Recently, possible violations to the Tan relation have been found in the presence of spin-orbit coupling \cite{Qin2020}, particle losses \cite{Bouchoule2021}, impurity-bath interactions in an expanding gas \cite{Cayla2022} and hard-wall boundaries \cite{Aupetit-Diallo2023}.~Tan law has been considered valid at temperatures $T$ even well above the critical value $T_c$ of the superfluid phase transition \cite{Hu2011, Braaten2012, Pitaevskii2016}, in contrast with the Maxwell-Boltzmann Gaussian decay expected in the classical gas limit.~Tan relation has been experimentally confirmed only at $T < T_c$ \cite{Stewart2010}, raising the question above which temperature it may be \textit{violated} \cite{Makotyn2014}. 

Atomic, solid-state, electronic and spin systems exhibit an \textit{Anomaly}, i.e., ~a thermal feature in the thermodynamic properties as a function of temperature, identified by a peak in the specific heat, a maximum in the chemical potential or a minimum in the magnetization, located at the anomaly temperature $T_A$ \cite{DeRosi2021II}.~The onset of a thermal second-order phase transition is signalled by an anomaly where $T_A = T_c$ \cite{Landau2013}.~In absence of a phase transition, the anomaly is due to unpopulated states in the excitation spectrum \cite{Raju1992,Harris1998,Tari2003,He2009,Lucas2017,Brambleby2017,Jurcisinova2018}.~When the temperature is comparable to $T_A$, empty spectral states are thermally occupied, the excitations experience the breakdown of the low-$T$ quasiparticle description \cite{Yan2020}, and thermal fluctuations dominate over quantum correlations at $T > T_A$ \cite{DeRosi2021II}.~Thus, the internal energy is almost constant with temperature at $T < T_A$, and rapidly increases at $T > T_A$ \cite{Ku2012}.~Anomalies are present \cite{DeRosi2021II} in any system in one spatial dimension (1D) \cite{Dender1997, Hammar1999, Nakanishi2002, Ruegg2008, Bouillot2011} where phase transitions are forbidden \cite{Landau2013}.  

In a 1D repulsive Bose gas, the \textit{Hole Anomaly} has been recently predicted for any contact interaction strength \cite{DeRosi2021II}.~This mechanism occurs through the thermal occupation of states located below the spectral hole branch whose maximum provides the energy scale for the anomaly temperature $T_A$.~Tan relation is confirmed by comparison with Path Integral Monte Carlo (PIMC) results at $T < T_A$ \cite{Xu2015}.~No knowledge at $T > T_A$ was available so far and an open question is how the tail of the momentum distribution $n\left(k\right)$ changes across $T_A$.  

In this work, we report that the thermal increase of the internal energy, induced by the hole anomaly, makes dominant the $1/|k|^3$-term, by screening the Tan relation $n\left(k\right) \sim \mathcal{C}/k^4$ at $T > T_A$.~This thermal fading occurs for any interaction strength at high temperatures as shown by PIMC results.~It may be observed in 1D atomic Bose gases where $n\left(k\right)$ has been measured \cite{Richard2003, vanAmerongen2008, Meinert2015, Yang2017} and the exploration of a wide range of interaction strength and temperature values is possible \cite{Salces-Carcoba2018}.

\lettersection{Model}

We consider a 1D uniform gas composed of $N$ Bose particles interacting via contact-pairwise repulsive potential and described by the Hamiltonian \cite{Mistakidis2023}:
\begin{equation}
\label{Eq:H}
H = - \frac{\hbar^2}{2m}\sum_{i = 1}^N \frac{\partial^2}{\partial x_i^2} + g\sum_{i > j}^N\delta(x_i - x_j),
\end{equation}
where $m$ is the particle mass, $g = - 2 \hbar^2/(m a)>0$ is the 1D coupling constant \cite{Olshanii1998}, and $a<0$ is the 1D $s$-wave scattering length.~We study the thermodynamic limit $N \to \infty$ by increasing the system size $L\to\infty$ while keeping the linear density $n=N/L$ fixed.~The interaction strength is $\gamma = - 2/(n a)$.

In 1D, there are no phase transitions but rather a continuous crossover that encompasses different regimes in terms of $\gamma$ and temperature.~The gas admits a mean-field description \cite{Pitaevskii2016} in the Gross-Pitaevskii limit of weak repulsion $\gamma \ll 1$, which in 1D corresponds to high density $n |a| \gg 1$.~In the Tonks-Girardeau regime of infinite repulsion $\gamma \to \infty$, achieved at low density $n |a| \to 0$, bosons become impenetrable and the wavefunction is mapped \cite{Girardeau1960} onto that of an ideal (non-interacting) Fermi gas, resulting in identical thermodynamics and spectrum.~Many experiments explored this interaction crossover in ultracold atom platforms \cite{Paredes2004, Kinoshita2004, Tolra2004, Kinoshita2005, Haller2011, Jacqmin2011, Guarrera2012}.~At zero temperature, the energetic properties can be obtained using the exact Bethe-Ansatz method \cite{Lieb1963, Lieb1963II, Pitaevskii2016, DeRosi2017}: the ground-state energy $E_0$, chemical potential $\mu_0 = (\partial E_0/\partial N)_{a, L}$ and speed of sound $v = \sqrt{n/m (\partial \mu_0/\partial n)_{a}}$ which are all functions of $\gamma$.

At finite temperature $T$, the exact thermal Bethe Ansatz (TBA) approach \cite{Yang1969, Yang1970} can be used and the thermodynamics within the canonical ensemble is captured by the Helmholtz free energy $A = E - TS$, where $E$ is the internal energy and $S$ the entropy.~The Tan contact can be obtained via \cite{Braaten2011, Yao2018, DeRosi2019} 
\begin{equation}
\label{Eq:contact}
\mathcal{C}= \left(4 m/ \hbar^2\right) \left(\partial A/\partial a\right)_{T,N,L}  ,
\end{equation}
which provides information on the interaction energy and a relation between the pressure and $E$ \cite{Olshanii2003, Tan2008, Tan2008II, Tan2008III, Barth2011, DeRosi2019, DeRosi2022}. 

\begin{figure}[h!]
\includegraphics[width=0.47\textwidth]{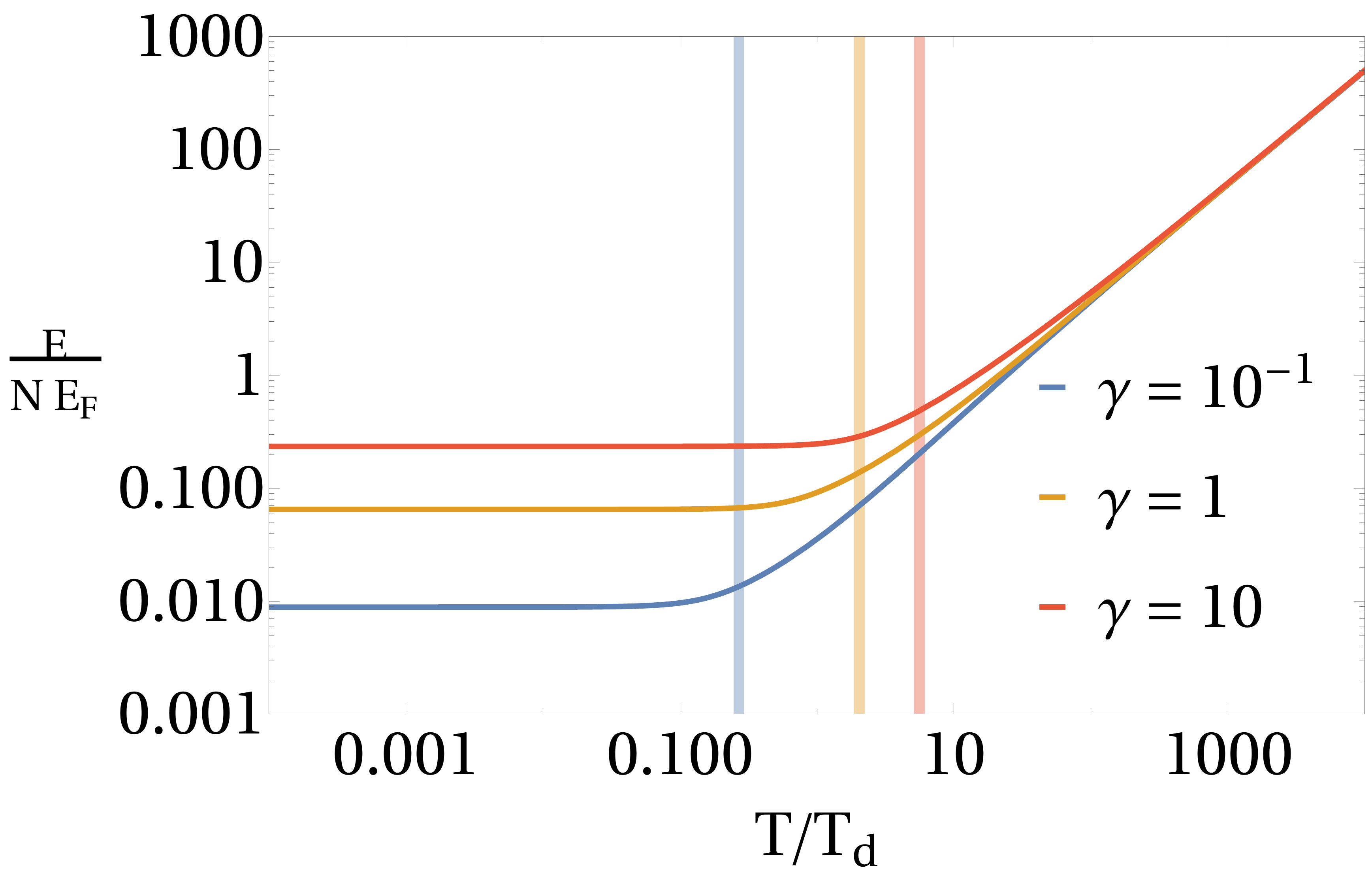}
\caption{Internal energy per particle $E/N$, normalized to the Fermi value $E_F = k_B T_F = \hbar^2 \pi^2 n^2/\left(2 m\right)$, vs temperature in units of the quantum degeneracy threshold $T_d = T_F/\pi^2$ and for several interaction strengths $\gamma$ reported from small (bottom) to large (top) values. Calculations have been performed with TBA. Vertical lines denote the anomaly temperature $T_A/T_d$ from small (left) to large (right) $\gamma$, corresponding to $0.27 $ ($\gamma = 10^{-1}$), $2.05$ ($\gamma = 1$) and $5.58$ ($\gamma = 10$).
}
\label{fig:Energy}
\end{figure}

In Fig.~\ref{fig:Energy}, we report the exact thermal Bethe-Ansatz results of the internal energy per particle $E/N$ as a function of temperature and for characteristic values of the interaction strength $\gamma$.~We show energies in units of the Fermi value $E_F = k_B T_F = \hbar^2 \pi^2 n^2/\left(2 m\right)$ and temperatures rescaled by the quantum degeneracy threshold $T_d = T_F/\pi^2$.~Vertical lines denote the hole-anomaly temperature $T_A/T_d$ estimated from the peak in the specific heat \cite{DeRosi2021II}.~For any $\gamma$, $E/N$ is almost constant at $T \lesssim T_A$, while it exhibits an intense monotonic increase at $T > T_A$, due to the thermal occupation of spectral states which is completed around $T_A$ \cite{DeRosi2021II}.

\begin{figure*}
\includegraphics[width=\textwidth]{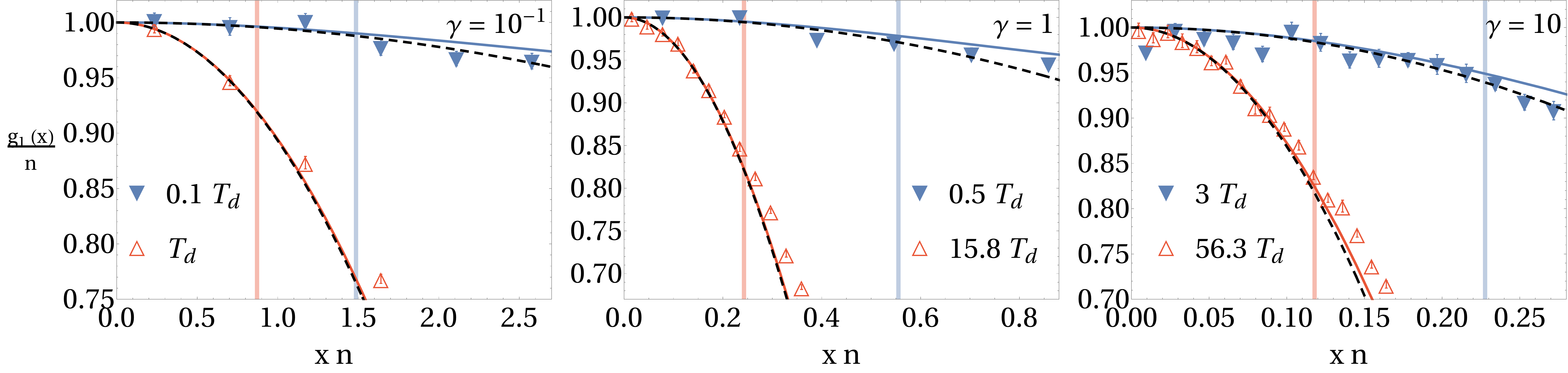}
\caption{OBDM $g_1\left(x\right)/n$ vs interparticle distance $x n$ ($n$ is the density) for the interaction strength $\gamma = 10^{-1}$ (first panel), $\gamma = 1$ (second), and $\gamma = 10$ (last).~Symbols denote PIMC results and their sizes are larger than the statistical error bars.~Solid (empty) symbols correspond to temperatures below (above) the anomaly value $T_A$, Fig.~\ref{fig:Energy}.~Solid lines represent the short-distance expansion \eqref{Eq:OBDM small x} calculated with TBA.~Dashed black lines correspond to Eq.~\eqref{Eq:OBDM small x} with $b_3 = 0$.~Curves are reported from low (top) to high (bottom) temperatures in each panel.~The maximal distance of the expansion $x_{\rm max} n$ \eqref{Eq:xmax} is shown with vertical lines from low (right) to high (left) temperatures at fixed $\gamma$.}
\label{fig:OBDM}
\end{figure*}

\lettersection{One-Body Density Matrix}

The one-body density matrix (OBDM) is defined as the non-diagonal density \cite{Pitaevskii2016}:  
 \begin{equation}
 \label{Eq:OBDM}
 g_1\left (x = x_1 - x_2 \right) = \langle  \hat{\psi}^\dagger\left(x_1\right)  \hat{\psi}\left(x_2\right) \rangle  ,
 \end{equation}
where $\hat{\psi}\left(x \right)$ is the Bose field, $x$ the interparticle distance, and $\langle\cdots\rangle$ the average over an ensemble in thermal equilibrium.~The OBDM quantifies the coherence and corresponds to the amplitude of the process where a particle is annihilated at position $x_2$ and another one is created at $x_1$.~At $x = 0$, one recovers the diagonal density $n$.~The momentum distribution is the Fourier transform of the OBDM.

We employ the PIMC method to calculate the complete $x$-dependence of the OBDM for a wide range of interaction strength $\gamma$ and temperatures in a 1D Bose gas \cite{DeRosi2022}.~At high temperatures, $T \gg T_d$, PIMC results show an excellent agreement with the Maxwell-Boltzmann (MB) Gaussian law $g_1\left(x \right)_{\rm MB}  = n e^{-x^2/(2\sigma^2)}$ \cite{DeRosi2022} describing a classical gas and decaying to zero for $x \gg \sigma$
where $\sigma = \lambda/\sqrt{2 \pi} $ is the standard deviation proportional to the thermal de Broglie wavelength $\lambda$.

The short-distance expansion of the OBDM is \cite{Astrakharchik2006II} 
\begin{equation}
\label{Eq:OBDM small x}
\frac{g_1\left(|x| \lesssim x_{\rm max} \right)}{n} = 1 + \sum_{i = 1}^\infty c_i \left(x n\right)^{i} + b_3 \left|x n\right|^3 + \mathcal{O}\left(|x n|^4\right). 
\end{equation}
The coefficients $c_i$ in the Taylor expansion of the analytic part are the corresponding moments of the momentum distribution \cite{Astrakharchik2006II}, they diverge for $i > 3$ and the odd ones vanish $c_1 = c_3 = \cdots = 0$.~From the Hellmann-Feynman theorem \cite{Feynman1939}, one finds that the second coefficient is a function of the internal energy $E/N$ and Tan contact $\mathcal{C}/N$ per particle:
\begin{equation}
\label{Eq:c2}
c_2 = - \frac{1}{2} \left( \frac{E}{N} \frac{2 m}{\hbar^2 n^2} - \frac{\mathcal{C}}{N }\frac{1}{\gamma n^3}  \right) ,
\end{equation}
and $c_2$ can be also expressed in terms of the average kinetic energy \cite{Pitaevskii2016, DeRosi2022}.~The non-analytic part of Eq.~\eqref{Eq:OBDM small x} starts with a $|x|^3$ dependence whose coefficient depends on $\mathcal{C}/N$ only:
\begin{equation}
\label{Eq:c3}
b_3 = \left(\mathcal{C}/N\right)/\left(12 n^3\right) \ .
\end{equation} 
At $T = 0$, Eqs.~(\ref{Eq:OBDM small x}-\ref{Eq:c3}) have been derived \cite{Olshanii2003} also including the coefficient of the $|x|^4$ term \cite{Yurovsky2008, Dunjko2011, Olshanii2017}.~The finite-temperature dependence enters in $E$ and $\mathcal{C}$ which can be evaluated with exact TBA \cite{Yang1969, Yang1970, DeRosi2019, DeRosi2022}. Eq.~\eqref{Eq:OBDM small x} is valid for any value of $\gamma$ and $T$ as shown by comparison with PIMC calculations \cite{DeRosi2022}. 

We find in this work that \eqref{Eq:OBDM small x} holds up to a maximal distance
\begin{equation}
\label{Eq:xmax}
x_{\rm max} = \left(\xi^{-1} + \sigma^{-1}\right)^{-1}
\end{equation}
which is determined by the healing length $\xi = \hbar/(\sqrt{2} m v)$ and the standard deviation $\sigma$ of the Gaussian $g_1\left(x \right)_{\rm MB}$.~At $T = 0$, the expression \eqref{Eq:xmax} reduces to $x_{\rm max} = \xi$ which constrains the high-momentum range for the tail of the momentum distribution $n\left(|k| \gtrsim \xi^{-1} \right)$ for any interaction strength, as shown previously by comparison with exact Monte Carlo results \cite{Cazalilla2004}.~Equation \eqref{Eq:xmax} holds in any system where the sound velocity $v$, entering in $\xi$ and depending on the interaction strength \cite{Pitaevskii2016}, is well-defined.~At very high temperatures, where the system approaches the Maxwell-Boltzmann regime, we recover the classical ideal gas limit $x_{\rm max} = \sigma$.~Equation \eqref{Eq:xmax} is a smooth interpolation between zero- and high-temperature limits, given by $\xi$ and $\sigma$, respectively.~Equation \eqref{Eq:xmax} provides an excellent approximation, for any interaction strength and temperature, of the threshold where the short-distance expansion \eqref{Eq:OBDM small x} deviates from the exact PIMC results for the OBDM, as discussed below.

In Fig.~\ref{fig:OBDM}, we show exact Path Integral Monte Carlo results of the one-body density matrix.~Solid symbols correspond to temperatures below the hole anomaly $T < T_A$, while empty ones for $T > T_A$, see Fig.~\ref{fig:Energy}.~Weakly- ($\gamma = 10^{-1}$, first panel), intermediate- ($\gamma = 1$, second) and strongly- ($\gamma = 10$, last) interacting regimes are reported.~We test the importance of the non-analytic contribution by comparing the short-distance OBDM, Eq.~\eqref{Eq:OBDM small x}, with $b_3 \neq 0$ (colored solid lines) and $b_3 = 0$ (black dashed), calculated with thermal Bethe-Ansatz.~The maximal distance $x_{\rm max}$ \eqref{Eq:xmax} is shown with vertical lines.

Our results are valid for any interaction strength $\gamma$.~The short-range expansion \eqref{Eq:OBDM small x} of the OBDM holds at distances limited by the upper bound~\eqref{Eq:xmax} ($|x| \lesssim x_{\rm max}$) at any temperature \footnote{
See Supplemental Material at [...] for additional results of the one-body density matrix and the momentum distribution, including the Maxwell-Boltzmann regime of the classical ideal gas at high temperatures. The Supplemental Material includes Refs.~\cite{Cazalilla2004, Esteve2006, DeRosi2022}.
}, as witnessed by the comparison with PIMC findings.~The non-analytic term (with coefficient $b_3$) plays a role in \eqref{Eq:OBDM small x} at distances close to $x_{\rm max}$ for an accurate description at $T < T_A$, while it is negligible at $T > T_A$ \cite{Note1}.~The reason is that the internal energy increases at $T > T_A$, see Fig.~\ref{fig:Energy}, making $b_3$ \eqref{Eq:c3} small compared to $c_2$~\eqref{Eq:c2}.~The thermal fading of the $|x|^3$-dependence in the short-distance OBDM is then driven by the hole anomaly.~It is a crossover \cite{Note1} and not an abrupt change which is expected crossing the critical temperature of a second-order phase transition.  

In the high-temperature MB regime~\cite{Note1},~we obtain $x_{\rm max} = \sigma$; $E/N = k_B T/2$ as the energy is defined by thermal fluctuations \cite{DeRosi2021II} and contact $\mathcal{C} = 0$ as interactions are negligible, Eq.~\eqref{Eq:contact}.~The short-range behavior~\eqref{Eq:OBDM small x} of the OBDM recovers the analytic terms of the expansion of the Gaussian $g_1\left(x \lesssim \sigma\right)_{\rm MB}/n = 1 - \left( x/\sigma \right)^2/2 + \mathcal{O}\left(x^4\right)$, where the non-analytic one is absent.~Even though $b_3$ \eqref{Eq:c3} can be omitted at $T > T_A$, $\mathcal{C}$ also enters into $c_2$ \eqref{Eq:c2} and still plays a role until very high temperatures are reached, where $g_1\left(x\right)_{\rm MB}$ is valid.

\lettersection{Momentum Distribution}

The momentum distribution is related to the OBDM~\eqref{Eq:OBDM} by a Fourier transform \cite{Pitaevskii2016}
\begin{equation}
\label{Eq:n(k)}
n\left(k\right) = \frac{1}{n} \int_{-\infty}^{+\infty} \frac{d x}{2 \pi \hbar} \cos\left(k \cdot x\right) g_1(x) \ ,
\end{equation}
and gives the probability to find an atom with momentum $k$. 

In a 1D Bose gas, $n\left(k\right)$ has been calculated at $T = 0$ with the diffusion Monte Carlo technique \cite{Astrakharchik2003, Astrakharchik2006}.~At finite temperature, various numerical and analytical methods have been applied but restricted to strong \cite{Drummond2004, Xu2015} and weak \cite{Mora2003} interactions, and temperatures below the hole anomaly \cite{Mora2003, Cheng2022}.

Our work fills an important gap:~we compute $n\left(k\right)$ in a 1D Bose gas with the most advanced PIMC method, exploring all regimes from weak to strong interactions and from low to high temperatures \cite{Note1}.~To this aim, we apply the Fourier transform \eqref{Eq:n(k)} to the PIMC results for the OBDM \cite{DeRosi2022}.~At high $T$, our PIMC data are captured by the Gaussian $n\left(k\right)_{\rm MB} = \sigma/\left( \hbar \sqrt{2 \pi} \right) e^{- \sigma^2 k^2/2}$ typical of a MB classical gas \cite{Note1}.

\begin{figure}
\includegraphics[width=0.47\textwidth]{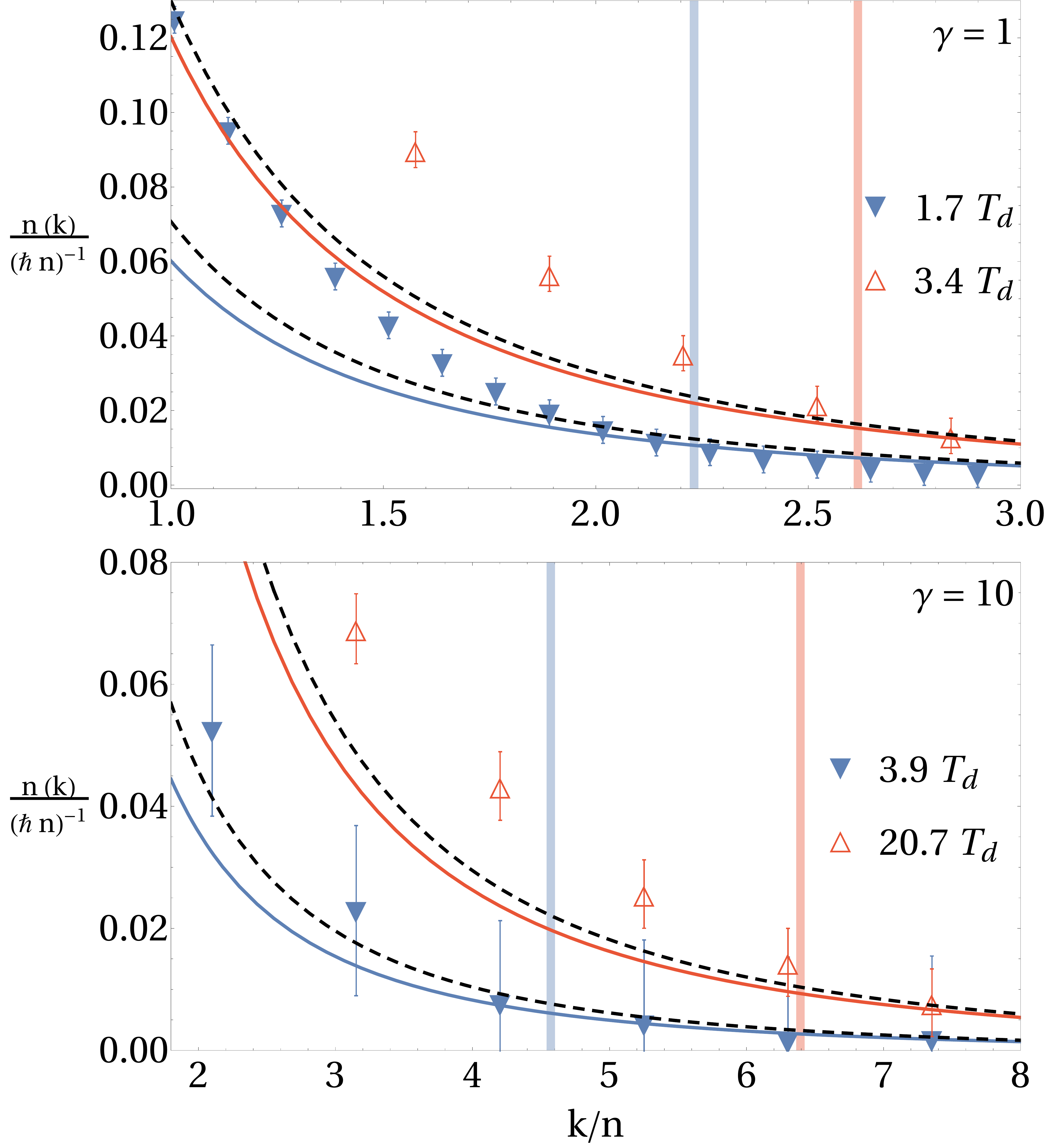}
\caption{Momentum distribution $n\left(k\right)$ vs momentum $k$ for $\gamma = 1$ (upper panel) and $\gamma = 10$ (lower panel).~Symbols correspond to PIMC results for $T < T_A$ (solid) and $T > T_A$ (empty).~Solid lines represent the tail \eqref{Eq:FT of short distance} calculated with TBA.~Dashed lines present \eqref{Eq:FT of short distance} with $b_3 = 0$.~Curves are reported from low (bottom) to high (top) temperatures in each panel.~$k_{\rm min} = x_{\rm max}^{-1}$ \eqref{Eq:xmax} is denoted by vertical lines from low (left) to high (right) temperatures at fixed $\gamma$.    
}
\label{fig:nkTail}
\end{figure}

\begin{figure}
\includegraphics[width=0.475\textwidth]{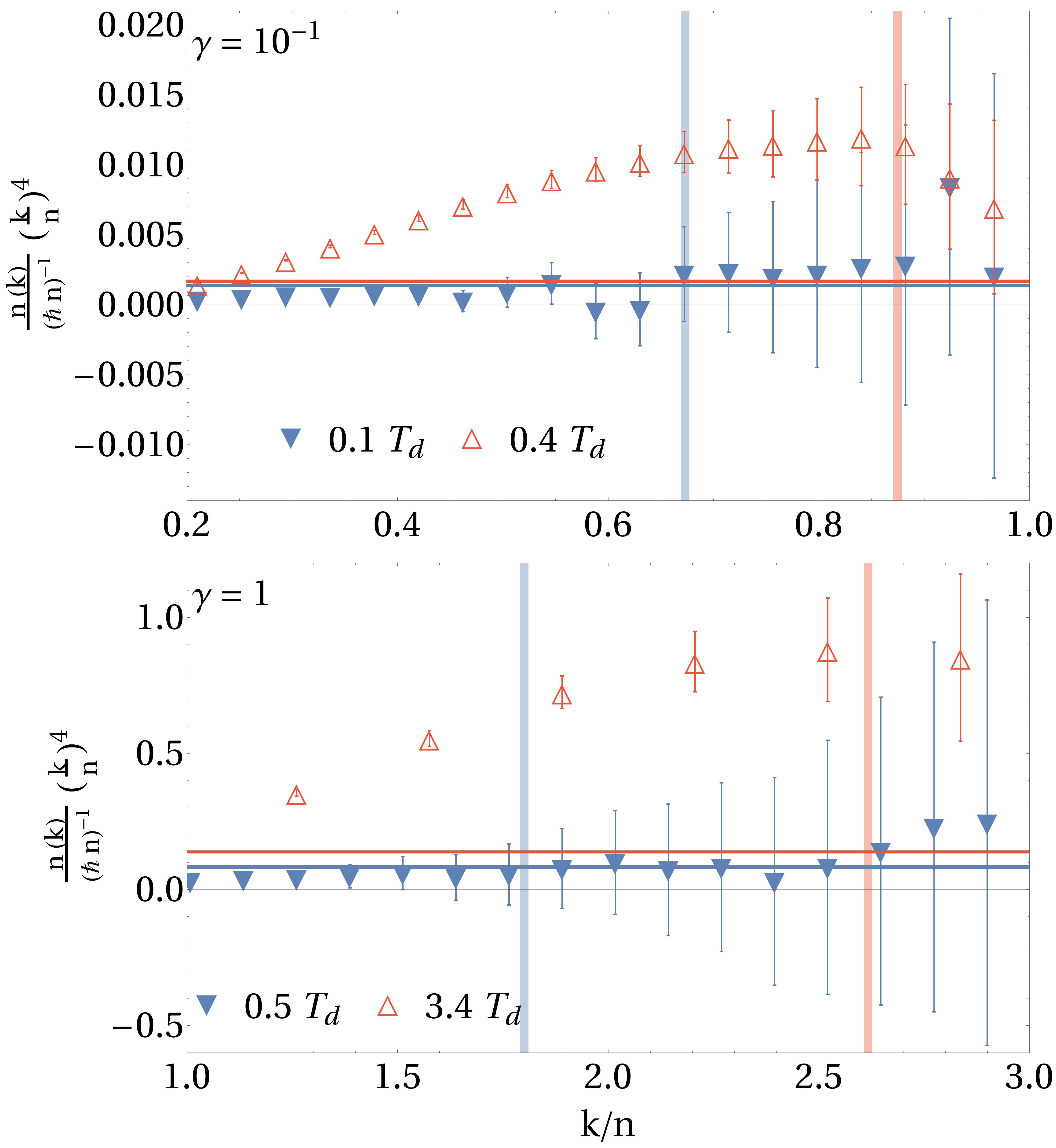}
\caption{Scaled momentum distribution $n\left(k\right) k^4$ for $\gamma = 10^{-1}$ ($\gamma = 1$) in the upper (lower) panel.~Solid (empty) symbols correspond to $T < T_A$ ($T > T_A$) and represent PIMC results.~$k_{\rm min}$ is denoted by vertical lines from low (left) to high (right) temperatures.~Horizontal lines report $6 n^3 b_3/\left(\pi \hbar \right)$ of \eqref{Eq:FT of short distance} obtained with TBA. 
}
\label{fig:nk4k}
\end{figure}

We derive the large-$k$ tail of $n\left(k\right)$ by using the short-distance OBDM~\eqref{Eq:OBDM small x} in Eq.~\eqref{Eq:n(k)}, where we integrate up to $x_{\rm max} = k_{\rm min}^{-1}$ \eqref{Eq:xmax} fixing the minimum momentum for the tail:
\begin{multline}
\label{Eq:FT of short distance}
n(|k| \gtrsim k_{\rm min}) =  \frac{6 n^3}{\pi \hbar} \frac{b_3}{k^4} \left[1 - \cos\left( \frac{k}{k_{\rm min}}  \right)     \right] \\
- \frac{1}{\pi \hbar|k|^3} \sin\left(\frac{|k|}{k_{\rm min}}\right)  \left( 2 c_2 n^2 + \frac{6 n^3 b_3}{ k_{\rm min}}  \right) + \mathcal{O}\left(\frac{1}{k^2} \right). 
\end{multline}
The $1/k^4$-term emerges from the leading non-analytic behavior of the short-distance OBDM and provides the Tan relation $\sim \mathcal{C}/k^4$, which, consistently with coordinate space, is more important at momenta closer to the lower bound $|k| \gtrsim k_{\rm min}$.~Tan relation has   
been derived for the 1D Bose gas at $T = 0$ \cite{Olshanii2003}. 
The $1/|k|^3$-contribution depends even on the $c_2$ coefficient, which is a function of the internal energy and contact \eqref{Eq:c2}, as well as the momentum $k_{\rm min}$. Equation~\eqref{Eq:FT of short distance} recovers the classical limit of the Fourier transform of $g_1\left(x \lesssim \sigma\right)_{\rm MB}$.

In Figs.~\ref{fig:nkTail}-\ref{fig:nk4k}, we show the correlations at large momenta and crossing the hole-anomaly temperature $T_A$ for several interaction strengths $\gamma$.~Symbols denote Path Integral Monte Carlo results for $T < T_A$ (solid) and $T > T_A$ (empty), see Fig.~\ref{fig:Energy}.~The minimal momentum $k_{\rm min} = x_{\rm max}^{-1}$ \eqref{Eq:xmax} for the tail \eqref{Eq:FT of short distance} is reported with vertical lines.~In Fig.~\ref{fig:nkTail}, solid and black dashed lines correspond to the tail \eqref{Eq:FT of short distance} with and without the non-analytic term, respectively, and are calculated with Thermal Bethe-Ansatz.~Fig.~\ref{fig:nk4k} presents $n\left(k\right) k^4$, horizontal lines denote the coefficient of the Tan relation, from which the deviation of PIMC predictions is evident at $T > T_A$.      

Consistently with our results for the short-distance OBDM, in Fig.~\ref{fig:nkTail}, the complete tail for the distribution \eqref{Eq:FT of short distance} and its minimal momentum $k_{\rm min}$ are excellent approximations for any interaction strength and temperature \cite{Note1}, as shown by comparison with PIMC findings.~While the Tan relation $\sim \mathcal{C}/k^4$ in \eqref{Eq:FT of short distance} is verified at low temperatures, it is thermally faded above the anomaly $T > T_A$.~This fading crossover appears for any interaction strength \cite{Note1} and at momenta close to $k_{\rm min}$, see Fig.~\ref{fig:nk4k}, where the Tan relation is more important in \eqref{Eq:FT of short distance}.~The deviation from the Tan law in Fig.~\ref{fig:nk4k} gets larger by raising the temperature \cite{Note1}.~However, the fading starts to occur at the anomaly temperature which is much lower than the one needed for the achievement of the Maxwell-Boltzmann classical gas regime, which is a limit described by Eq.~\eqref{Eq:FT of short distance}.

\lettersection{Experimental Considerations}

1D atomic Bose gases can be realized with a single optical tube trap \cite{Salces-Carcoba2018} which allows for the exploration of temperatures below and above the hole anomaly.~Spatial uniform density is achieved with a flatbox potential \cite{Rauer2018}.~The interaction strength $\gamma \sim 1/\left(n a\right)$ is tuned by changing the density $n$ via the strongly confining radial potential \cite{Olshanii1998, Kinoshita2004, Kinoshita2005} or by adjusting the scattering length $a$ through Fano-Feshbach resonances \cite{Chin2010, Meinert2015}.~The temperature can be extracted from a single absorption image during time-of-flight expansion with neural network \cite{Moller2021}.~The momentum distribution can also be measured \cite{Richard2003, Paredes2004, vanAmerongen2008, Fabbri2011, Meinert2015, Yang2017, Wilson2020, Malvania2021, Le2023}. 

\lettersection{Conclusions}

We built a unified theory which describes the entire contact-interaction and temperature crossover in a 1D repulsive Bose gas and provides a novel connection between excitations and correlations.~We report that the hole anomaly, due to the thermal occupation of spectral states, induces an increase of the internal energy which is responsible for the high-temperature fading of the Tan relation in the large-momentum (short-distance) one-body correlations.~Anomalies are ubiquitous in a variety of systems \cite{DeRosi2021II}, even with interactions beyond the $s$-wave pairwise contact model, and behave as a second-order phase transition at the critical temperature.~Our work suggests that the anomaly temperature may be identified in many systems \cite{Astrakharchik2005, Astrakharchik2006II, Haller2009, Stewart2010, Fischer2014, DeRosi2017II, DeRosi2021, Hofmann2021, Li2023, DelMaestro2022} with the change from the quantum to thermal behavior in correlations even at short and not only at large \cite{Fischer2014, Murthy2015} distances.

\begin{acknowledgments} 
G.~D.~R. received funding from the grant IJC2020-043542-I funded by MCIN/AEI/10.13039/501100011033 and by "European Union NextGenerationEU/PRTR".~G.D.R., G.E.A. and J.B. were partially supported by the grant PID2020-113565GB-C21 funded by MCIN/AEI/10.13039/501100011033 and the grant 2021 SGR 01411 from the Generalitat de Catalunya.~M.O. was supported by the NSF grant No. PHY-1912542.
\end{acknowledgments}

\bibliography{Bibliography}

\begin{thebibliography}{91}%
\makeatletter
\providecommand \@ifxundefined [1]{%
 \@ifx{#1\undefined}
}%
\providecommand \@ifnum [1]{%
 \ifnum #1\expandafter \@firstoftwo
 \else \expandafter \@secondoftwo
 \fi
}%
\providecommand \@ifx [1]{%
 \ifx #1\expandafter \@firstoftwo
 \else \expandafter \@secondoftwo
 \fi
}%
\providecommand \natexlab [1]{#1}%
\providecommand \enquote  [1]{``#1''}%
\providecommand \bibnamefont  [1]{#1}%
\providecommand \bibfnamefont [1]{#1}%
\providecommand \citenamefont [1]{#1}%
\providecommand \href@noop [0]{\@secondoftwo}%
\providecommand \href [0]{\begingroup \@sanitize@url \@href}%
\providecommand \@href[1]{\@@startlink{#1}\@@href}%
\providecommand \@@href[1]{\endgroup#1\@@endlink}%
\providecommand \@sanitize@url [0]{\catcode `\\12\catcode `\$12\catcode
  `\&12\catcode `\#12\catcode `\^12\catcode `\_12\catcode `\%12\relax}%
\providecommand \@@startlink[1]{}%
\providecommand \@@endlink[0]{}%
\providecommand \url  [0]{\begingroup\@sanitize@url \@url }%
\providecommand \@url [1]{\endgroup\@href {#1}{\urlprefix }}%
\providecommand \urlprefix  [0]{URL }%
\providecommand \Eprint [0]{\href }%
\providecommand \doibase [0]{http://dx.doi.org/}%
\providecommand \selectlanguage [0]{\@gobble}%
\providecommand \bibinfo  [0]{\@secondoftwo}%
\providecommand \bibfield  [0]{\@secondoftwo}%
\providecommand \translation [1]{[#1]}%
\providecommand \BibitemOpen [0]{}%
\providecommand \bibitemStop [0]{}%
\providecommand \bibitemNoStop [0]{.\EOS\space}%
\providecommand \EOS [0]{\spacefactor3000\relax}%
\providecommand \BibitemShut  [1]{\csname bibitem#1\endcsname}%
\let\auto@bib@innerbib\@empty
\bibitem [{\citenamefont {Tan}(2008{\natexlab{a}})}]{Tan2008}%
  \BibitemOpen
  \bibfield  {author} {\bibinfo {author} {\bibfnamefont {S.}~\bibnamefont
  {Tan}},\ }\bibfield  {title} {\bibinfo {title} {\emph {Energetics of a
  strongly correlated Fermi gas}},\ }\href {\doibase
  https://doi.org/10.1016/j.aop.2008.03.004} {\bibfield  {journal} {\bibinfo
  {journal} {Annals of Physics}\ }\textbf {\bibinfo {volume} {323}},\ \bibinfo
  {pages} {2952 } (\bibinfo {year} {2008}{\natexlab{a}})}\BibitemShut {NoStop}%
\bibitem [{\citenamefont {Tan}(2008{\natexlab{b}})}]{Tan2008II}%
  \BibitemOpen
  \bibfield  {author} {\bibinfo {author} {\bibfnamefont {S.}~\bibnamefont
  {Tan}},\ }\bibfield  {title} {\bibinfo {title} {\emph {Generalized virial
  theorem and pressure relation for a strongly correlated Fermi gas}},\ }\href
  {\doibase https://doi.org/10.1016/j.aop.2008.03.003} {\bibfield  {journal}
  {\bibinfo  {journal} {Annals of Physics}\ }\textbf {\bibinfo {volume}
  {323}},\ \bibinfo {pages} {2987 } (\bibinfo {year}
  {2008}{\natexlab{b}})}\BibitemShut {NoStop}%
\bibitem [{\citenamefont {Tan}(2008{\natexlab{c}})}]{Tan2008III}%
  \BibitemOpen
  \bibfield  {author} {\bibinfo {author} {\bibfnamefont {S.}~\bibnamefont
  {Tan}},\ }\bibfield  {title} {\bibinfo {title} {\emph {Large momentum part of
  a strongly correlated Fermi gas}},\ }\href {\doibase
  https://doi.org/10.1016/j.aop.2008.03.005} {\bibfield  {journal} {\bibinfo
  {journal} {Annals of Physics}\ }\textbf {\bibinfo {volume} {323}},\ \bibinfo
  {pages} {2971 } (\bibinfo {year} {2008}{\natexlab{c}})}\BibitemShut {NoStop}%
\bibitem [{\citenamefont {Bulgac}(2023)}]{Bulgac2023}%
  \BibitemOpen
  \bibfield  {author} {\bibinfo {author} {\bibfnamefont {A.}~\bibnamefont
  {Bulgac}},\ }\bibfield  {title} {\bibinfo {title} {\emph {Entanglement
  entropy, single-particle occupation probabilities, and short-range
  correlations}},\ }\href {\doibase 10.1103/PhysRevC.107.L061602} {\bibfield
  {journal} {\bibinfo  {journal} {Phys. Rev. C}\ }\textbf {\bibinfo {volume}
  {107}},\ \bibinfo {pages} {L061602} (\bibinfo {year} {2023})}\BibitemShut
  {NoStop}%
\bibitem [{\citenamefont {Olshanii}\ and\ \citenamefont
  {Dunjko}(2003)}]{Olshanii2003}%
  \BibitemOpen
  \bibfield  {author} {\bibinfo {author} {\bibfnamefont {M.}~\bibnamefont
  {Olshanii}}\ and\ \bibinfo {author} {\bibfnamefont {V.}~\bibnamefont
  {Dunjko}},\ }\bibfield  {title} {\bibinfo {title} {\emph {Short-Distance
  Correlation Properties of the Lieb-Liniger System and Momentum Distributions
  of Trapped One-Dimensional Atomic Gases}},\ }\href {\doibase
  10.1103/PhysRevLett.91.090401} {\bibfield  {journal} {\bibinfo  {journal}
  {Phys. Rev. Lett.}\ }\textbf {\bibinfo {volume} {91}},\ \bibinfo {pages}
  {090401} (\bibinfo {year} {2003})}\BibitemShut {NoStop}%
\bibitem [{\citenamefont {Barth}\ and\ \citenamefont
  {Zwerger}(2011)}]{Barth2011}%
  \BibitemOpen
  \bibfield  {author} {\bibinfo {author} {\bibfnamefont {M.}~\bibnamefont
  {Barth}}\ and\ \bibinfo {author} {\bibfnamefont {W.}~\bibnamefont
  {Zwerger}},\ }\bibfield  {title} {\bibinfo {title} {\emph {Tan relations in
  one dimension}},\ }\href {\doibase https://doi.org/10.1016/j.aop.2011.05.010}
  {\bibfield  {journal} {\bibinfo  {journal} {Annals of Physics}\ }\textbf
  {\bibinfo {volume} {326}},\ \bibinfo {pages} {2544 } (\bibinfo {year}
  {2011})}\BibitemShut {NoStop}%
\bibitem [{\citenamefont {Matveeva}\ and\ \citenamefont
  {Astrakharchik}(2016)}]{Matveeva_2016}%
  \BibitemOpen
  \bibfield  {author} {\bibinfo {author} {\bibfnamefont {N.}~\bibnamefont
  {Matveeva}}\ and\ \bibinfo {author} {\bibfnamefont {G.~E.}\ \bibnamefont
  {Astrakharchik}},\ }\bibfield  {title} {\bibinfo {title} {\emph
  {One-dimensional multicomponent Fermi gas in a trap: quantum Monte Carlo
  study}},\ }\href {\doibase 10.1088/1367-2630/18/6/065009} {\bibfield
  {journal} {\bibinfo  {journal} {New Journal of Physics}\ }\textbf {\bibinfo
  {volume} {18}},\ \bibinfo {pages} {065009} (\bibinfo {year}
  {2016})}\BibitemShut {NoStop}%
\bibitem [{\citenamefont {P\^a\ifmmode~\mbox{\c{t}}\else \c{t}\fi{}u}\ and\
  \citenamefont {Kl\"umper}(2017)}]{Patu2017}%
  \BibitemOpen
  \bibfield  {author} {\bibinfo {author} {\bibfnamefont {O.~I.}\ \bibnamefont
  {P\^a\ifmmode~\mbox{\c{t}}\else \c{t}\fi{}u}}\ and\ \bibinfo {author}
  {\bibfnamefont {A.}~\bibnamefont {Kl\"umper}},\ }\bibfield  {title} {\bibinfo
  {title} {\emph {Universal Tan relations for quantum gases in one
  dimension}},\ }\href {\doibase 10.1103/PhysRevA.96.063612} {\bibfield
  {journal} {\bibinfo  {journal} {Phys. Rev. A}\ }\textbf {\bibinfo {volume}
  {96}},\ \bibinfo {pages} {063612} (\bibinfo {year} {2017})}\BibitemShut
  {NoStop}%
\bibitem [{\citenamefont {Minguzzi}\ \emph {et~al.}(2002)\citenamefont
  {Minguzzi}, \citenamefont {Vignolo},\ and\ \citenamefont
  {Tosi}}]{Minguzzi2002}%
  \BibitemOpen
  \bibfield  {author} {\bibinfo {author} {\bibfnamefont {A.}~\bibnamefont
  {Minguzzi}}, \bibinfo {author} {\bibfnamefont {P.}~\bibnamefont {Vignolo}}, \
  and\ \bibinfo {author} {\bibfnamefont {M.~P.}\ \bibnamefont {Tosi}},\
  }\bibfield  {title} {\bibinfo {title} {\emph {High-momentum tail in the Tonks
  gas under harmonic confinement}},\ }\href {\doibase
  10.1016/S0375-9601(02)00042-7} {\bibfield  {journal} {\bibinfo  {journal}
  {Physics Letters A}\ }\textbf {\bibinfo {volume} {294}},\ \bibinfo {pages}
  {222} (\bibinfo {year} {2002})}\BibitemShut {NoStop}%
\bibitem [{\citenamefont {Rigol}(2005)}]{Rigol2005}%
  \BibitemOpen
  \bibfield  {author} {\bibinfo {author} {\bibfnamefont {M.}~\bibnamefont
  {Rigol}},\ }\bibfield  {title} {\bibinfo {title} {\emph {Finite-temperature
  properties of hard-core bosons confined on one-dimensional optical
  lattices}},\ }\href {\doibase 10.1103/PhysRevA.72.063607} {\bibfield
  {journal} {\bibinfo  {journal} {Phys. Rev. A}\ }\textbf {\bibinfo {volume}
  {72}},\ \bibinfo {pages} {063607} (\bibinfo {year} {2005})}\BibitemShut
  {NoStop}%
\bibitem [{\citenamefont {Werner}\ and\ \citenamefont
  {Castin}(2012{\natexlab{a}})}]{Werner2012}%
  \BibitemOpen
  \bibfield  {author} {\bibinfo {author} {\bibfnamefont {F.}~\bibnamefont
  {Werner}}\ and\ \bibinfo {author} {\bibfnamefont {Y.}~\bibnamefont
  {Castin}},\ }\bibfield  {title} {\bibinfo {title} {\emph {General relations
  for quantum gases in two and three dimensions: Two-component fermions}},\
  }\href {\doibase 10.1103/PhysRevA.86.013626} {\bibfield  {journal} {\bibinfo
  {journal} {Phys. Rev. A}\ }\textbf {\bibinfo {volume} {86}},\ \bibinfo
  {pages} {013626} (\bibinfo {year} {2012}{\natexlab{a}})}\BibitemShut
  {NoStop}%
\bibitem [{\citenamefont {Werner}\ and\ \citenamefont
  {Castin}(2012{\natexlab{b}})}]{Werner2012b}%
  \BibitemOpen
  \bibfield  {author} {\bibinfo {author} {\bibfnamefont {F.}~\bibnamefont
  {Werner}}\ and\ \bibinfo {author} {\bibfnamefont {Y.}~\bibnamefont
  {Castin}},\ }\bibfield  {title} {\bibinfo {title} {\emph {General relations
  for quantum gases in two and three dimensions. II. Bosons and mixtures}},\
  }\href {\doibase 10.1103/PhysRevA.86.053633} {\bibfield  {journal} {\bibinfo
  {journal} {Phys. Rev. A}\ }\textbf {\bibinfo {volume} {86}},\ \bibinfo
  {pages} {053633} (\bibinfo {year} {2012}{\natexlab{b}})}\BibitemShut
  {NoStop}%
\bibitem [{\citenamefont {Braaten}(2012)}]{Braaten2012}%
  \BibitemOpen
  \bibfield  {author} {\bibinfo {author} {\bibfnamefont {E.}~\bibnamefont
  {Braaten}},\ }\bibinfo {title} {\emph {Universal Relations for Fermions with
  Large Scattering Length}},\ in\ \href {\doibase 10.1007/978-3-642-21978-8_6}
  {\emph {\bibinfo {booktitle} {The BCS-BEC Crossover and the Unitary Fermi
  Gas}}},\ \bibinfo {editor} {edited by\ \bibinfo {editor} {\bibfnamefont
  {W.}~\bibnamefont {Zwerger}}}\ (\bibinfo  {publisher} {Springer Berlin
  Heidelberg},\ \bibinfo {address} {Berlin, Heidelberg},\ \bibinfo {year}
  {2012})\ pp.\ \bibinfo {pages} {193--231}\BibitemShut {NoStop}%
\bibitem [{\citenamefont {Pitaevskii}\ and\ \citenamefont
  {Stringari}(2016)}]{Pitaevskii2016}%
  \BibitemOpen
  \bibfield  {author} {\bibinfo {author} {\bibfnamefont {L.}~\bibnamefont
  {Pitaevskii}}\ and\ \bibinfo {author} {\bibfnamefont {S.}~\bibnamefont
  {Stringari}},\ }\href@noop {} {\emph {\bibinfo {title} {Bose-Einstein
  Condensation and Superfluidity}}},\ International Series of Monographs on
  Physics\ (\bibinfo  {publisher} {Oxford University Press, Oxford},\ \bibinfo
  {year} {2016})\BibitemShut {NoStop}%
\bibitem [{\citenamefont {Qin}\ and\ \citenamefont {Zhang}(2020)}]{Qin2020}%
  \BibitemOpen
  \bibfield  {author} {\bibinfo {author} {\bibfnamefont {F.}~\bibnamefont
  {Qin}}\ and\ \bibinfo {author} {\bibfnamefont {P.}~\bibnamefont {Zhang}},\
  }\bibfield  {title} {\bibinfo {title} {\emph {Universal relations for
  hybridized $s$- and $p$-wave interactions from spin-orbital coupling}},\
  }\href {\doibase 10.1103/PhysRevA.102.043321} {\bibfield  {journal} {\bibinfo
   {journal} {Phys. Rev. A}\ }\textbf {\bibinfo {volume} {102}},\ \bibinfo
  {pages} {043321} (\bibinfo {year} {2020})}\BibitemShut {NoStop}%
\bibitem [{\citenamefont {Bouchoule}\ and\ \citenamefont
  {Dubail}(2021)}]{Bouchoule2021}%
  \BibitemOpen
  \bibfield  {author} {\bibinfo {author} {\bibfnamefont {I.}~\bibnamefont
  {Bouchoule}}\ and\ \bibinfo {author} {\bibfnamefont {J.}~\bibnamefont
  {Dubail}},\ }\bibfield  {title} {\bibinfo {title} {\emph {Breakdown of Tan's
  Relation in Lossy One-Dimensional Bose Gases}},\ }\href {\doibase
  10.1103/PhysRevLett.126.160603} {\bibfield  {journal} {\bibinfo  {journal}
  {Phys. Rev. Lett.}\ }\textbf {\bibinfo {volume} {126}},\ \bibinfo {pages}
  {160603} (\bibinfo {year} {2021})}\BibitemShut {NoStop}%
\bibitem [{\citenamefont {Cayla}\ \emph {et~al.}(2023)\citenamefont {Cayla},
  \citenamefont {Massignan}, \citenamefont {Giamarchi}, \citenamefont {Aspect},
  \citenamefont {Westbrook},\ and\ \citenamefont {Cl\'ement}}]{Cayla2022}%
  \BibitemOpen
  \bibfield  {author} {\bibinfo {author} {\bibfnamefont {H.}~\bibnamefont
  {Cayla}}, \bibinfo {author} {\bibfnamefont {P.}~\bibnamefont {Massignan}},
  \bibinfo {author} {\bibfnamefont {T.}~\bibnamefont {Giamarchi}}, \bibinfo
  {author} {\bibfnamefont {A.}~\bibnamefont {Aspect}}, \bibinfo {author}
  {\bibfnamefont {C.~I.}\ \bibnamefont {Westbrook}}, \ and\ \bibinfo {author}
  {\bibfnamefont {D.}~\bibnamefont {Cl\'ement}},\ }\bibfield  {title} {\bibinfo
  {title} {\emph {Observation of $1/{k}^{4}$-Tails after Expansion of
  Bose-Einstein Condensates with Impurities}},\ }\href {\doibase
  10.1103/PhysRevLett.130.153401} {\bibfield  {journal} {\bibinfo  {journal}
  {Phys. Rev. Lett.}\ }\textbf {\bibinfo {volume} {130}},\ \bibinfo {pages}
  {153401} (\bibinfo {year} {2023})}\BibitemShut {NoStop}%
\bibitem [{\citenamefont {Aupetit-Diallo}\ \emph {et~al.}(2023)\citenamefont
  {Aupetit-Diallo}, \citenamefont {Musolino}, \citenamefont {Albert},\ and\
  \citenamefont {Vignolo}}]{Aupetit-Diallo2023}%
  \BibitemOpen
  \bibfield  {author} {\bibinfo {author} {\bibfnamefont {G.}~\bibnamefont
  {Aupetit-Diallo}}, \bibinfo {author} {\bibfnamefont {S.}~\bibnamefont
  {Musolino}}, \bibinfo {author} {\bibfnamefont {M.}~\bibnamefont {Albert}}, \
  and\ \bibinfo {author} {\bibfnamefont {P.}~\bibnamefont {Vignolo}},\
  }\bibfield  {title} {\bibinfo {title} {\emph {High-momentum oscillating tails
  of strongly interacting one-dimensional gases in a box}},\ }\href {\doibase
  10.1103/PhysRevA.107.L061301} {\bibfield  {journal} {\bibinfo  {journal}
  {Phys. Rev. A}\ }\textbf {\bibinfo {volume} {107}},\ \bibinfo {pages}
  {L061301} (\bibinfo {year} {2023})}\BibitemShut {NoStop}%
\bibitem [{\citenamefont {Hu}\ \emph {et~al.}(2011)\citenamefont {Hu},
  \citenamefont {Liu},\ and\ \citenamefont {Drummond}}]{Hu2011}%
  \BibitemOpen
  \bibfield  {author} {\bibinfo {author} {\bibfnamefont {H.}~\bibnamefont
  {Hu}}, \bibinfo {author} {\bibfnamefont {X.-J.}\ \bibnamefont {Liu}}, \ and\
  \bibinfo {author} {\bibfnamefont {P.~D.}\ \bibnamefont {Drummond}},\
  }\bibfield  {title} {\bibinfo {title} {\emph {Universal contact of strongly
  interacting fermions at finite temperatures}},\ }\href {\doibase
  10.1088/1367-2630/13/3/035007} {\bibfield  {journal} {\bibinfo  {journal}
  {New Journal of Physics}\ }\textbf {\bibinfo {volume} {13}},\ \bibinfo
  {pages} {035007} (\bibinfo {year} {2011})}\BibitemShut {NoStop}%
\bibitem [{\citenamefont {Stewart}\ \emph {et~al.}(2010)\citenamefont
  {Stewart}, \citenamefont {Gaebler}, \citenamefont {Drake},\ and\
  \citenamefont {Jin}}]{Stewart2010}%
  \BibitemOpen
  \bibfield  {author} {\bibinfo {author} {\bibfnamefont {J.~T.}\ \bibnamefont
  {Stewart}}, \bibinfo {author} {\bibfnamefont {J.~P.}\ \bibnamefont
  {Gaebler}}, \bibinfo {author} {\bibfnamefont {T.~E.}\ \bibnamefont {Drake}},
  \ and\ \bibinfo {author} {\bibfnamefont {D.~S.}\ \bibnamefont {Jin}},\
  }\bibfield  {title} {\bibinfo {title} {\emph {Verification of Universal
  Relations in a Strongly Interacting Fermi Gas}},\ }\href {\doibase
  10.1103/PhysRevLett.104.235301} {\bibfield  {journal} {\bibinfo  {journal}
  {Phys. Rev. Lett.}\ }\textbf {\bibinfo {volume} {104}},\ \bibinfo {pages}
  {235301} (\bibinfo {year} {2010})}\BibitemShut {NoStop}%
\bibitem [{\citenamefont {Makotyn}\ \emph {et~al.}(2014)\citenamefont
  {Makotyn}, \citenamefont {Klauss}, \citenamefont {Goldberger}, \citenamefont
  {Cornell},\ and\ \citenamefont {Jin}}]{Makotyn2014}%
  \BibitemOpen
  \bibfield  {author} {\bibinfo {author} {\bibfnamefont {P.}~\bibnamefont
  {Makotyn}}, \bibinfo {author} {\bibfnamefont {C.~E.}\ \bibnamefont {Klauss}},
  \bibinfo {author} {\bibfnamefont {D.~L.}\ \bibnamefont {Goldberger}},
  \bibinfo {author} {\bibfnamefont {E.~A.}\ \bibnamefont {Cornell}}, \ and\
  \bibinfo {author} {\bibfnamefont {D.~S.}\ \bibnamefont {Jin}},\ }\bibfield
  {title} {\bibinfo {title} {\emph {Universal dynamics of a degenerate unitary
  Bose gas}},\ }\href {\doibase 10.1038/nphys2850} {\bibfield  {journal}
  {\bibinfo  {journal} {Nature Physics}\ }\textbf {\bibinfo {volume} {10}},\
  \bibinfo {pages} {116} (\bibinfo {year} {2014})}\BibitemShut {NoStop}%
\bibitem [{\citenamefont {De~Rosi}\ \emph {et~al.}(2022)\citenamefont
  {De~Rosi}, \citenamefont {Rota}, \citenamefont {Astrakharchik},\ and\
  \citenamefont {Boronat}}]{DeRosi2021II}%
  \BibitemOpen
  \bibfield  {author} {\bibinfo {author} {\bibfnamefont {G.}~\bibnamefont
  {De~Rosi}}, \bibinfo {author} {\bibfnamefont {R.}~\bibnamefont {Rota}},
  \bibinfo {author} {\bibfnamefont {G.~E.}\ \bibnamefont {Astrakharchik}}, \
  and\ \bibinfo {author} {\bibfnamefont {J.}~\bibnamefont {Boronat}},\
  }\bibfield  {title} {\bibinfo {title} {\emph {{Hole-induced anomaly in the
  thermodynamic behavior of a one-dimensional Bose gas}}},\ }\href {\doibase
  10.21468/SciPostPhys.13.2.035} {\bibfield  {journal} {\bibinfo  {journal}
  {SciPost Phys.}\ }\textbf {\bibinfo {volume} {13}},\ \bibinfo {pages} {035}
  (\bibinfo {year} {2022})}\BibitemShut {NoStop}%
\bibitem [{\citenamefont {Landau}\ and\ \citenamefont
  {Lifshitz}(2013)}]{Landau2013}%
  \BibitemOpen
  \bibfield  {author} {\bibinfo {author} {\bibfnamefont {L.~D.}\ \bibnamefont
  {Landau}}\ and\ \bibinfo {author} {\bibfnamefont {E.~M.}\ \bibnamefont
  {Lifshitz}},\ }\href@noop {} {\emph {\bibinfo {title} {Statistical Physics:
  Vol. 5}}}\ (\bibinfo  {publisher} {Elsevier Science},\ \bibinfo {year}
  {2013})\BibitemShut {NoStop}%
\bibitem [{\citenamefont {Raju}\ \emph {et~al.}(1992)\citenamefont {Raju},
  \citenamefont {Gmelin},\ and\ \citenamefont {Kremer}}]{Raju1992}%
  \BibitemOpen
  \bibfield  {author} {\bibinfo {author} {\bibfnamefont {N.~P.}\ \bibnamefont
  {Raju}}, \bibinfo {author} {\bibfnamefont {E.}~\bibnamefont {Gmelin}}, \ and\
  \bibinfo {author} {\bibfnamefont {R.~K.}\ \bibnamefont {Kremer}},\ }\bibfield
   {title} {\bibinfo {title} {\emph {Magnetic-susceptibility and specific-heat
  studies of spin-glass-like ordering in the pyrochlore compounds
  ${\mathit{R}}_{2}$${\mathrm{Mo}}_{2}$${\mathrm{O}}_{7}$ (R=Y, Sm, or Gd)}},\
  }\href {\doibase 10.1103/PhysRevB.46.5405} {\bibfield  {journal} {\bibinfo
  {journal} {Phys. Rev. B}\ }\textbf {\bibinfo {volume} {46}},\ \bibinfo
  {pages} {5405} (\bibinfo {year} {1992})}\BibitemShut {NoStop}%
\bibitem [{\citenamefont {Harris}\ \emph {et~al.}(1998)\citenamefont {Harris},
  \citenamefont {Bramwell}, \citenamefont {Holdsworth},\ and\ \citenamefont
  {Champion}}]{Harris1998}%
  \BibitemOpen
  \bibfield  {author} {\bibinfo {author} {\bibfnamefont {M.~J.}\ \bibnamefont
  {Harris}}, \bibinfo {author} {\bibfnamefont {S.~T.}\ \bibnamefont
  {Bramwell}}, \bibinfo {author} {\bibfnamefont {P.~C.~W.}\ \bibnamefont
  {Holdsworth}}, \ and\ \bibinfo {author} {\bibfnamefont {J.~D.~M.}\
  \bibnamefont {Champion}},\ }\bibfield  {title} {\bibinfo {title} {\emph
  {Liquid-Gas Critical Behavior in a Frustrated Pyrochlore Ferromagnet}},\
  }\href {\doibase 10.1103/PhysRevLett.81.4496} {\bibfield  {journal} {\bibinfo
   {journal} {Phys. Rev. Lett.}\ }\textbf {\bibinfo {volume} {81}},\ \bibinfo
  {pages} {4496} (\bibinfo {year} {1998})}\BibitemShut {NoStop}%
\bibitem [{\citenamefont {Tari}(2003)}]{Tari2003}%
  \BibitemOpen
  \bibfield  {author} {\bibinfo {author} {\bibfnamefont {A.}~\bibnamefont
  {Tari}},\ }\href@noop {} {\emph {\bibinfo {title} {The Specific Heat Of
  Matter At Low Temperatures}}}\ (\bibinfo  {publisher} {World Scientific
  Publishing Company},\ \bibinfo {year} {2003})\BibitemShut {NoStop}%
\bibitem [{\citenamefont {He}\ \emph {et~al.}(2009)\citenamefont {He},
  \citenamefont {Zheng}, \citenamefont {Mitchell}, \citenamefont {Foo},
  \citenamefont {Cava},\ and\ \citenamefont {Leighton}}]{He2009}%
  \BibitemOpen
  \bibfield  {author} {\bibinfo {author} {\bibfnamefont {C.}~\bibnamefont
  {He}}, \bibinfo {author} {\bibfnamefont {H.}~\bibnamefont {Zheng}}, \bibinfo
  {author} {\bibfnamefont {J.~F.}\ \bibnamefont {Mitchell}}, \bibinfo {author}
  {\bibfnamefont {M.~L.}\ \bibnamefont {Foo}}, \bibinfo {author} {\bibfnamefont
  {R.~J.}\ \bibnamefont {Cava}}, \ and\ \bibinfo {author} {\bibfnamefont
  {C.}~\bibnamefont {Leighton}},\ }\bibfield  {title} {\bibinfo {title} {\emph
  {Low temperature Schottky anomalies in the specific heat of LaCoO3:
  Defect-stabilized finite spin states}},\ }\href {\doibase 10.1063/1.3098374}
  {\bibfield  {journal} {\bibinfo  {journal} {Applied Physics Letters}\
  }\textbf {\bibinfo {volume} {94}},\ \bibinfo {pages} {102514} (\bibinfo
  {year} {2009})}\BibitemShut {NoStop}%
\bibitem [{\citenamefont {Lucas}\ \emph {et~al.}(2017)\citenamefont {Lucas},
  \citenamefont {Grube}, \citenamefont {Huang}, \citenamefont {Sakai},
  \citenamefont {Wunderlich}, \citenamefont {Green}, \citenamefont {Wosnitza},
  \citenamefont {Fritsch}, \citenamefont {Gegenwart}, \citenamefont
  {Stockert},\ and\ \citenamefont {v.~L\"ohneysen}}]{Lucas2017}%
  \BibitemOpen
  \bibfield  {author} {\bibinfo {author} {\bibfnamefont {S.}~\bibnamefont
  {Lucas}}, \bibinfo {author} {\bibfnamefont {K.}~\bibnamefont {Grube}},
  \bibinfo {author} {\bibfnamefont {C.-L.}\ \bibnamefont {Huang}}, \bibinfo
  {author} {\bibfnamefont {A.}~\bibnamefont {Sakai}}, \bibinfo {author}
  {\bibfnamefont {S.}~\bibnamefont {Wunderlich}}, \bibinfo {author}
  {\bibfnamefont {E.~L.}\ \bibnamefont {Green}}, \bibinfo {author}
  {\bibfnamefont {J.}~\bibnamefont {Wosnitza}}, \bibinfo {author}
  {\bibfnamefont {V.}~\bibnamefont {Fritsch}}, \bibinfo {author} {\bibfnamefont
  {P.}~\bibnamefont {Gegenwart}}, \bibinfo {author} {\bibfnamefont
  {O.}~\bibnamefont {Stockert}}, \ and\ \bibinfo {author} {\bibfnamefont
  {H.}~\bibnamefont {v.~L\"ohneysen}},\ }\bibfield  {title} {\bibinfo {title}
  {\emph {Entropy Evolution in the Magnetic Phases of Partially Frustrated
  CePdAl}},\ }\href {\doibase 10.1103/PhysRevLett.118.107204} {\bibfield
  {journal} {\bibinfo  {journal} {Phys. Rev. Lett.}\ }\textbf {\bibinfo
  {volume} {118}},\ \bibinfo {pages} {107204} (\bibinfo {year}
  {2017})}\BibitemShut {NoStop}%
\bibitem [{\citenamefont {Brambleby}\ \emph {et~al.}(2017)\citenamefont
  {Brambleby}, \citenamefont {Goddard}, \citenamefont {Singleton},
  \citenamefont {Jaime}, \citenamefont {Lancaster}, \citenamefont {Huang},
  \citenamefont {Wosnitza}, \citenamefont {Topping}, \citenamefont {Carreiro},
  \citenamefont {Tran}, \citenamefont {Manson},\ and\ \citenamefont
  {Manson}}]{Brambleby2017}%
  \BibitemOpen
  \bibfield  {author} {\bibinfo {author} {\bibfnamefont {J.}~\bibnamefont
  {Brambleby}}, \bibinfo {author} {\bibfnamefont {P.~A.}\ \bibnamefont
  {Goddard}}, \bibinfo {author} {\bibfnamefont {J.}~\bibnamefont {Singleton}},
  \bibinfo {author} {\bibfnamefont {M.}~\bibnamefont {Jaime}}, \bibinfo
  {author} {\bibfnamefont {T.}~\bibnamefont {Lancaster}}, \bibinfo {author}
  {\bibfnamefont {L.}~\bibnamefont {Huang}}, \bibinfo {author} {\bibfnamefont
  {J.}~\bibnamefont {Wosnitza}}, \bibinfo {author} {\bibfnamefont {C.~V.}\
  \bibnamefont {Topping}}, \bibinfo {author} {\bibfnamefont {K.~E.}\
  \bibnamefont {Carreiro}}, \bibinfo {author} {\bibfnamefont {H.~E.}\
  \bibnamefont {Tran}}, \bibinfo {author} {\bibfnamefont {Z.~E.}\ \bibnamefont
  {Manson}}, \ and\ \bibinfo {author} {\bibfnamefont {J.~L.}\ \bibnamefont
  {Manson}},\ }\bibfield  {title} {\bibinfo {title} {\emph {Adiabatic physics
  of an exchange-coupled spin-dimer system: Magnetocaloric effect, zero-point
  fluctuations, and possible two-dimensional universal behavior}},\ }\href
  {\doibase 10.1103/PhysRevB.95.024404} {\bibfield  {journal} {\bibinfo
  {journal} {Phys. Rev. B}\ }\textbf {\bibinfo {volume} {95}},\ \bibinfo
  {pages} {024404} (\bibinfo {year} {2017})}\BibitemShut {NoStop}%
\bibitem [{\citenamefont {Jur\ifmmode \check{c}\else
  \v{c}\fi{}i\ifmmode~\check{s}\else \v{s}\fi{}inov\'a}\ and\ \citenamefont
  {Jur\ifmmode \check{c}\else \v{c}\fi{}i\ifmmode~\check{s}\else
  \v{s}\fi{}in}(2018)}]{Jurcisinova2018}%
  \BibitemOpen
  \bibfield  {author} {\bibinfo {author} {\bibfnamefont {E.}~\bibnamefont
  {Jur\ifmmode \check{c}\else \v{c}\fi{}i\ifmmode~\check{s}\else
  \v{s}\fi{}inov\'a}}\ and\ \bibinfo {author} {\bibfnamefont {M.}~\bibnamefont
  {Jur\ifmmode \check{c}\else \v{c}\fi{}i\ifmmode~\check{s}\else
  \v{s}\fi{}in}},\ }\bibfield  {title} {\bibinfo {title} {\emph {Multipeak
  low-temperature behavior of specific heat capacity in frustrated magnetic
  systems: An exact theoretical analysis}},\ }\href {\doibase
  10.1103/PhysRevE.97.052129} {\bibfield  {journal} {\bibinfo  {journal} {Phys.
  Rev. E}\ }\textbf {\bibinfo {volume} {97}},\ \bibinfo {pages} {052129}
  (\bibinfo {year} {2018})}\BibitemShut {NoStop}%
\bibitem [{\citenamefont {Yan}\ \emph {et~al.}(2020)\citenamefont {Yan},
  \citenamefont {Ni}, \citenamefont {Robens},\ and\ \citenamefont
  {Zwierlein}}]{Yan2020}%
  \BibitemOpen
  \bibfield  {author} {\bibinfo {author} {\bibfnamefont {Z.~Z.}\ \bibnamefont
  {Yan}}, \bibinfo {author} {\bibfnamefont {Y.}~\bibnamefont {Ni}}, \bibinfo
  {author} {\bibfnamefont {C.}~\bibnamefont {Robens}}, \ and\ \bibinfo {author}
  {\bibfnamefont {M.~W.}\ \bibnamefont {Zwierlein}},\ }\bibfield  {title}
  {\bibinfo {title} {\emph {Bose polarons near quantum criticality}},\ }\href
  {\doibase 10.1126/science.aax5850} {\bibfield  {journal} {\bibinfo  {journal}
  {Science}\ }\textbf {\bibinfo {volume} {368}},\ \bibinfo {pages} {190}
  (\bibinfo {year} {2020})}\BibitemShut {NoStop}%
\bibitem [{\citenamefont {Ku}\ \emph {et~al.}(2012)\citenamefont {Ku},
  \citenamefont {Sommer}, \citenamefont {Cheuk},\ and\ \citenamefont
  {Zwierlein}}]{Ku2012}%
  \BibitemOpen
  \bibfield  {author} {\bibinfo {author} {\bibfnamefont {M.~J.~H.}\
  \bibnamefont {Ku}}, \bibinfo {author} {\bibfnamefont {A.~T.}\ \bibnamefont
  {Sommer}}, \bibinfo {author} {\bibfnamefont {L.~W.}\ \bibnamefont {Cheuk}}, \
  and\ \bibinfo {author} {\bibfnamefont {M.~W.}\ \bibnamefont {Zwierlein}},\
  }\bibfield  {title} {\bibinfo {title} {\emph {Revealing the Superfluid Lambda
  Transition in the Universal Thermodynamics of a Unitary Fermi Gas}},\ }\href
  {\doibase 10.1126/science.1214987} {\bibfield  {journal} {\bibinfo  {journal}
  {Science}\ }\textbf {\bibinfo {volume} {335}},\ \bibinfo {pages} {563}
  (\bibinfo {year} {2012})}\BibitemShut {NoStop}%
\bibitem [{\citenamefont {Dender}\ \emph {et~al.}(1997)\citenamefont {Dender},
  \citenamefont {Hammar}, \citenamefont {Reich}, \citenamefont {Broholm},\ and\
  \citenamefont {Aeppli}}]{Dender1997}%
  \BibitemOpen
  \bibfield  {author} {\bibinfo {author} {\bibfnamefont {D.~C.}\ \bibnamefont
  {Dender}}, \bibinfo {author} {\bibfnamefont {P.~R.}\ \bibnamefont {Hammar}},
  \bibinfo {author} {\bibfnamefont {D.~H.}\ \bibnamefont {Reich}}, \bibinfo
  {author} {\bibfnamefont {C.}~\bibnamefont {Broholm}}, \ and\ \bibinfo
  {author} {\bibfnamefont {G.}~\bibnamefont {Aeppli}},\ }\bibfield  {title}
  {\bibinfo {title} {\emph {Direct Observation of Field-Induced Incommensurate
  Fluctuations in a One-Dimensional S = 1/2 Antiferromagnet}},\ }\href
  {\doibase 10.1103/PhysRevLett.79.1750} {\bibfield  {journal} {\bibinfo
  {journal} {Phys. Rev. Lett.}\ }\textbf {\bibinfo {volume} {79}},\ \bibinfo
  {pages} {1750 } (\bibinfo {year} {1997})}\BibitemShut {NoStop}%
\bibitem [{\citenamefont {Hammar}\ \emph {et~al.}(1999)\citenamefont {Hammar},
  \citenamefont {Stone}, \citenamefont {Reich}, \citenamefont {Broholm},
  \citenamefont {Gibson}, \citenamefont {Turnbull}, \citenamefont {Landee},\
  and\ \citenamefont {Oshikawa}}]{Hammar1999}%
  \BibitemOpen
  \bibfield  {author} {\bibinfo {author} {\bibfnamefont {P.~R.}\ \bibnamefont
  {Hammar}}, \bibinfo {author} {\bibfnamefont {M.~B.}\ \bibnamefont {Stone}},
  \bibinfo {author} {\bibfnamefont {D.~H.}\ \bibnamefont {Reich}}, \bibinfo
  {author} {\bibfnamefont {C.}~\bibnamefont {Broholm}}, \bibinfo {author}
  {\bibfnamefont {P.~J.}\ \bibnamefont {Gibson}}, \bibinfo {author}
  {\bibfnamefont {M.~M.}\ \bibnamefont {Turnbull}}, \bibinfo {author}
  {\bibfnamefont {C.~P.}\ \bibnamefont {Landee}}, \ and\ \bibinfo {author}
  {\bibfnamefont {M.}~\bibnamefont {Oshikawa}},\ }\bibfield  {title} {\bibinfo
  {title} {\emph {Characterization of a quasi-one-dimensional spin-1/2 magnet
  which is gapless and paramagnetic for
  $g{\ensuremath{\mu}}_{B}H\ensuremath{\lesssim}J$ and
  ${k}_{B}T\ensuremath{\ll}J$}},\ }\href {\doibase 10.1103/PhysRevB.59.1008}
  {\bibfield  {journal} {\bibinfo  {journal} {Phys. Rev. B}\ }\textbf {\bibinfo
  {volume} {59}},\ \bibinfo {pages} {1008} (\bibinfo {year}
  {1999})}\BibitemShut {NoStop}%
\bibitem [{\citenamefont {Nakanishi}\ and\ \citenamefont
  {Yamamoto}(2002)}]{Nakanishi2002}%
  \BibitemOpen
  \bibfield  {author} {\bibinfo {author} {\bibfnamefont {T.}~\bibnamefont
  {Nakanishi}}\ and\ \bibinfo {author} {\bibfnamefont {S.}~\bibnamefont
  {Yamamoto}},\ }\bibfield  {title} {\bibinfo {title} {\emph {Intrinsic
  double-peak structure of the specific heat in low-dimensional quantum
  ferrimagnets}},\ }\href {\doibase 10.1103/PhysRevB.65.214418} {\bibfield
  {journal} {\bibinfo  {journal} {Phys. Rev. B}\ }\textbf {\bibinfo {volume}
  {65}},\ \bibinfo {pages} {214418} (\bibinfo {year} {2002})}\BibitemShut
  {NoStop}%
\bibitem [{\citenamefont {R\"uegg}\ \emph {et~al.}(2008)\citenamefont
  {R\"uegg}, \citenamefont {Kiefer}, \citenamefont {Thielemann}, \citenamefont
  {McMorrow}, \citenamefont {Zapf}, \citenamefont {Normand}, \citenamefont
  {Zvonarev}, \citenamefont {Bouillot}, \citenamefont {Kollath}, \citenamefont
  {Giamarchi}, \citenamefont {Capponi}, \citenamefont {Poilblanc},
  \citenamefont {Biner},\ and\ \citenamefont {Kr\"amer}}]{Ruegg2008}%
  \BibitemOpen
  \bibfield  {author} {\bibinfo {author} {\bibfnamefont {C.}~\bibnamefont
  {R\"uegg}}, \bibinfo {author} {\bibfnamefont {K.}~\bibnamefont {Kiefer}},
  \bibinfo {author} {\bibfnamefont {B.}~\bibnamefont {Thielemann}}, \bibinfo
  {author} {\bibfnamefont {D.~F.}\ \bibnamefont {McMorrow}}, \bibinfo {author}
  {\bibfnamefont {V.}~\bibnamefont {Zapf}}, \bibinfo {author} {\bibfnamefont
  {B.}~\bibnamefont {Normand}}, \bibinfo {author} {\bibfnamefont {M.~B.}\
  \bibnamefont {Zvonarev}}, \bibinfo {author} {\bibfnamefont {P.}~\bibnamefont
  {Bouillot}}, \bibinfo {author} {\bibfnamefont {C.}~\bibnamefont {Kollath}},
  \bibinfo {author} {\bibfnamefont {T.}~\bibnamefont {Giamarchi}}, \bibinfo
  {author} {\bibfnamefont {S.}~\bibnamefont {Capponi}}, \bibinfo {author}
  {\bibfnamefont {D.}~\bibnamefont {Poilblanc}}, \bibinfo {author}
  {\bibfnamefont {D.}~\bibnamefont {Biner}}, \ and\ \bibinfo {author}
  {\bibfnamefont {K.~W.}\ \bibnamefont {Kr\"amer}},\ }\bibfield  {title}
  {\bibinfo {title} {\emph {Thermodynamics of the Spin Luttinger Liquid in a
  Model Ladder Material}},\ }\href {\doibase 10.1103/PhysRevLett.101.247202}
  {\bibfield  {journal} {\bibinfo  {journal} {Phys. Rev. Lett.}\ }\textbf
  {\bibinfo {volume} {101}},\ \bibinfo {pages} {247202} (\bibinfo {year}
  {2008})}\BibitemShut {NoStop}%
\bibitem [{\citenamefont {Bouillot}\ \emph {et~al.}(2011)\citenamefont
  {Bouillot}, \citenamefont {Kollath}, \citenamefont {L\"auchli}, \citenamefont
  {Zvonarev}, \citenamefont {Thielemann}, \citenamefont {R\"uegg},
  \citenamefont {Orignac}, \citenamefont {Citro}, \citenamefont
  {Klanj\ifmmode~\check{s}\else \v{s}\fi{}ek}, \citenamefont {Berthier},
  \citenamefont {Horvati\ifmmode~\acute{c}\else \'{c}\fi{}},\ and\
  \citenamefont {Giamarchi}}]{Bouillot2011}%
  \BibitemOpen
  \bibfield  {author} {\bibinfo {author} {\bibfnamefont {P.}~\bibnamefont
  {Bouillot}}, \bibinfo {author} {\bibfnamefont {C.}~\bibnamefont {Kollath}},
  \bibinfo {author} {\bibfnamefont {A.~M.}\ \bibnamefont {L\"auchli}}, \bibinfo
  {author} {\bibfnamefont {M.}~\bibnamefont {Zvonarev}}, \bibinfo {author}
  {\bibfnamefont {B.}~\bibnamefont {Thielemann}}, \bibinfo {author}
  {\bibfnamefont {C.}~\bibnamefont {R\"uegg}}, \bibinfo {author} {\bibfnamefont
  {E.}~\bibnamefont {Orignac}}, \bibinfo {author} {\bibfnamefont
  {R.}~\bibnamefont {Citro}}, \bibinfo {author} {\bibfnamefont
  {M.}~\bibnamefont {Klanj\ifmmode~\check{s}\else \v{s}\fi{}ek}}, \bibinfo
  {author} {\bibfnamefont {C.}~\bibnamefont {Berthier}}, \bibinfo {author}
  {\bibfnamefont {M.}~\bibnamefont {Horvati\ifmmode~\acute{c}\else
  \'{c}\fi{}}}, \ and\ \bibinfo {author} {\bibfnamefont {T.}~\bibnamefont
  {Giamarchi}},\ }\bibfield  {title} {\bibinfo {title} {\emph {Statics and
  dynamics of weakly coupled antiferromagnetic spin-$\frac{1}{2}$ ladders in a
  magnetic field}},\ }\href {\doibase 10.1103/PhysRevB.83.054407} {\bibfield
  {journal} {\bibinfo  {journal} {Phys. Rev. B}\ }\textbf {\bibinfo {volume}
  {83}},\ \bibinfo {pages} {054407} (\bibinfo {year} {2011})}\BibitemShut
  {NoStop}%
\bibitem [{\citenamefont {Xu}\ and\ \citenamefont {Rigol}(2015)}]{Xu2015}%
  \BibitemOpen
  \bibfield  {author} {\bibinfo {author} {\bibfnamefont {W.}~\bibnamefont
  {Xu}}\ and\ \bibinfo {author} {\bibfnamefont {M.}~\bibnamefont {Rigol}},\
  }\bibfield  {title} {\bibinfo {title} {\emph {Universal scaling of density
  and momentum distributions in Lieb-Liniger gases}},\ }\href {\doibase
  10.1103/PhysRevA.92.063623} {\bibfield  {journal} {\bibinfo  {journal} {Phys.
  Rev. A}\ }\textbf {\bibinfo {volume} {92}},\ \bibinfo {pages} {063623}
  (\bibinfo {year} {2015})}\BibitemShut {NoStop}%
\bibitem [{\citenamefont {Richard}\ \emph {et~al.}(2003)\citenamefont
  {Richard}, \citenamefont {Gerbier}, \citenamefont {Thywissen}, \citenamefont
  {Hugbart}, \citenamefont {Bouyer},\ and\ \citenamefont
  {Aspect}}]{Richard2003}%
  \BibitemOpen
  \bibfield  {author} {\bibinfo {author} {\bibfnamefont {S.}~\bibnamefont
  {Richard}}, \bibinfo {author} {\bibfnamefont {F.}~\bibnamefont {Gerbier}},
  \bibinfo {author} {\bibfnamefont {J.~H.}\ \bibnamefont {Thywissen}}, \bibinfo
  {author} {\bibfnamefont {M.}~\bibnamefont {Hugbart}}, \bibinfo {author}
  {\bibfnamefont {P.}~\bibnamefont {Bouyer}}, \ and\ \bibinfo {author}
  {\bibfnamefont {A.}~\bibnamefont {Aspect}},\ }\bibfield  {title} {\bibinfo
  {title} {\emph {Momentum Spectroscopy of 1D Phase Fluctuations in
  Bose-Einstein Condensates}},\ }\href {\doibase 10.1103/PhysRevLett.91.010405}
  {\bibfield  {journal} {\bibinfo  {journal} {Phys. Rev. Lett.}\ }\textbf
  {\bibinfo {volume} {91}},\ \bibinfo {pages} {010405} (\bibinfo {year}
  {2003})}\BibitemShut {NoStop}%
\bibitem [{\citenamefont {van Amerongen}\ \emph {et~al.}(2008)\citenamefont
  {van Amerongen}, \citenamefont {van Es}, \citenamefont {Wicke}, \citenamefont
  {Kheruntsyan},\ and\ \citenamefont {van Druten}}]{vanAmerongen2008}%
  \BibitemOpen
  \bibfield  {author} {\bibinfo {author} {\bibfnamefont {A.~H.}\ \bibnamefont
  {van Amerongen}}, \bibinfo {author} {\bibfnamefont {J.~J.~P.}\ \bibnamefont
  {van Es}}, \bibinfo {author} {\bibfnamefont {P.}~\bibnamefont {Wicke}},
  \bibinfo {author} {\bibfnamefont {K.~V.}\ \bibnamefont {Kheruntsyan}}, \ and\
  \bibinfo {author} {\bibfnamefont {N.~J.}\ \bibnamefont {van Druten}},\
  }\bibfield  {title} {\bibinfo {title} {\emph {Yang-Yang Thermodynamics on an
  Atom Chip}},\ }\href {\doibase 10.1103/PhysRevLett.100.090402} {\bibfield
  {journal} {\bibinfo  {journal} {Phys. Rev. Lett.}\ }\textbf {\bibinfo
  {volume} {100}},\ \bibinfo {pages} {090402} (\bibinfo {year}
  {2008})}\BibitemShut {NoStop}%
\bibitem [{\citenamefont {Meinert}\ \emph {et~al.}(2015)\citenamefont
  {Meinert}, \citenamefont {Panfil}, \citenamefont {Mark}, \citenamefont
  {Lauber}, \citenamefont {Caux},\ and\ \citenamefont
  {N\"agerl}}]{Meinert2015}%
  \BibitemOpen
  \bibfield  {author} {\bibinfo {author} {\bibfnamefont {F.}~\bibnamefont
  {Meinert}}, \bibinfo {author} {\bibfnamefont {M.}~\bibnamefont {Panfil}},
  \bibinfo {author} {\bibfnamefont {M.~J.}\ \bibnamefont {Mark}}, \bibinfo
  {author} {\bibfnamefont {K.}~\bibnamefont {Lauber}}, \bibinfo {author}
  {\bibfnamefont {J.-S.}\ \bibnamefont {Caux}}, \ and\ \bibinfo {author}
  {\bibfnamefont {H.-C.}\ \bibnamefont {N\"agerl}},\ }\bibfield  {title}
  {\bibinfo {title} {\emph {Probing the Excitations of a Lieb-Liniger Gas from
  Weak to Strong Coupling}},\ }\href {\doibase 10.1103/PhysRevLett.115.085301}
  {\bibfield  {journal} {\bibinfo  {journal} {Phys. Rev. Lett.}\ }\textbf
  {\bibinfo {volume} {115}},\ \bibinfo {pages} {085301} (\bibinfo {year}
  {2015})}\BibitemShut {NoStop}%
\bibitem [{\citenamefont {Yang}\ \emph {et~al.}(2017)\citenamefont {Yang},
  \citenamefont {Chen}, \citenamefont {Zheng}, \citenamefont {Sun},
  \citenamefont {Dai}, \citenamefont {Guan}, \citenamefont {Yuan},\ and\
  \citenamefont {Pan}}]{Yang2017}%
  \BibitemOpen
  \bibfield  {author} {\bibinfo {author} {\bibfnamefont {B.}~\bibnamefont
  {Yang}}, \bibinfo {author} {\bibfnamefont {Y.-Y.}\ \bibnamefont {Chen}},
  \bibinfo {author} {\bibfnamefont {Y.-G.}\ \bibnamefont {Zheng}}, \bibinfo
  {author} {\bibfnamefont {H.}~\bibnamefont {Sun}}, \bibinfo {author}
  {\bibfnamefont {H.-N.}\ \bibnamefont {Dai}}, \bibinfo {author} {\bibfnamefont
  {X.-W.}\ \bibnamefont {Guan}}, \bibinfo {author} {\bibfnamefont {Z.-S.}\
  \bibnamefont {Yuan}}, \ and\ \bibinfo {author} {\bibfnamefont {J.-W.}\
  \bibnamefont {Pan}},\ }\bibfield  {title} {\bibinfo {title} {\emph {Quantum
  criticality and the Tomonaga-Luttinger liquid in one-dimensional Bose
  gases}},\ }\href {\doibase 10.1103/PhysRevLett.119.165701} {\bibfield
  {journal} {\bibinfo  {journal} {Phys. Rev. Lett.}\ }\textbf {\bibinfo
  {volume} {119}},\ \bibinfo {pages} {165701} (\bibinfo {year}
  {2017})}\BibitemShut {NoStop}%
\bibitem [{\citenamefont {Salces-Carcoba}\ \emph {et~al.}(2018)\citenamefont
  {Salces-Carcoba}, \citenamefont {Billington}, \citenamefont {Putra},
  \citenamefont {Yue}, \citenamefont {Sugawa},\ and\ \citenamefont
  {Spielman}}]{Salces-Carcoba2018}%
  \BibitemOpen
  \bibfield  {author} {\bibinfo {author} {\bibfnamefont {F.}~\bibnamefont
  {Salces-Carcoba}}, \bibinfo {author} {\bibfnamefont {C.~J.}\ \bibnamefont
  {Billington}}, \bibinfo {author} {\bibfnamefont {A.}~\bibnamefont {Putra}},
  \bibinfo {author} {\bibfnamefont {Y.}~\bibnamefont {Yue}}, \bibinfo {author}
  {\bibfnamefont {S.}~\bibnamefont {Sugawa}}, \ and\ \bibinfo {author}
  {\bibfnamefont {I.~B.}\ \bibnamefont {Spielman}},\ }\bibfield  {title}
  {\bibinfo {title} {\emph {Equations of state from individual one-dimensional
  Bose gases}},\ }\href {http://stacks.iop.org/1367-2630/20/i=11/a=113032}
  {\bibfield  {journal} {\bibinfo  {journal} {New J. Phys.}\ }\textbf {\bibinfo
  {volume} {20}},\ \bibinfo {pages} {113032} (\bibinfo {year}
  {2018})}\BibitemShut {NoStop}%
\bibitem [{\citenamefont {Mistakidis}\ \emph {et~al.}(2023)\citenamefont
  {Mistakidis}, \citenamefont {Volosniev}, \citenamefont {Barfknecht},
  \citenamefont {Fogarty}, \citenamefont {Busch}, \citenamefont {Foerster},
  \citenamefont {Schmelcher},\ and\ \citenamefont {Zinner}}]{Mistakidis2023}%
  \BibitemOpen
  \bibfield  {author} {\bibinfo {author} {\bibfnamefont {S.}~\bibnamefont
  {Mistakidis}}, \bibinfo {author} {\bibfnamefont {A.}~\bibnamefont
  {Volosniev}}, \bibinfo {author} {\bibfnamefont {R.}~\bibnamefont
  {Barfknecht}}, \bibinfo {author} {\bibfnamefont {T.}~\bibnamefont {Fogarty}},
  \bibinfo {author} {\bibfnamefont {T.}~\bibnamefont {Busch}}, \bibinfo
  {author} {\bibfnamefont {A.}~\bibnamefont {Foerster}}, \bibinfo {author}
  {\bibfnamefont {P.}~\bibnamefont {Schmelcher}}, \ and\ \bibinfo {author}
  {\bibfnamefont {N.}~\bibnamefont {Zinner}},\ }\bibfield  {title} {\bibinfo
  {title} {\emph {Few-body Bose gases in low dimensions-A laboratory for
  quantum dynamics}},\ }\href {\doibase
  https://doi.org/10.1016/j.physrep.2023.10.004} {\bibfield  {journal}
  {\bibinfo  {journal} {Physics Reports}\ }\textbf {\bibinfo {volume} {1042}},\
  \bibinfo {pages} {1} (\bibinfo {year} {2023})}\BibitemShut {NoStop}%
\bibitem [{\citenamefont {Olshanii}(1998)}]{Olshanii1998}%
  \BibitemOpen
  \bibfield  {author} {\bibinfo {author} {\bibfnamefont {M.}~\bibnamefont
  {Olshanii}},\ }\bibfield  {title} {\bibinfo {title} {\emph {Atomic Scattering
  in the Presence of an External Confinement and a Gas of Impenetrable
  Bosons}},\ }\href {\doibase 10.1103/PhysRevLett.81.938} {\bibfield  {journal}
  {\bibinfo  {journal} {Phys. Rev. Lett.}\ }\textbf {\bibinfo {volume} {81}},\
  \bibinfo {pages} {938} (\bibinfo {year} {1998})}\BibitemShut {NoStop}%
\bibitem [{\citenamefont {Girardeau}(1960)}]{Girardeau1960}%
  \BibitemOpen
  \bibfield  {author} {\bibinfo {author} {\bibfnamefont {M.}~\bibnamefont
  {Girardeau}},\ }\bibfield  {title} {\bibinfo {title} {\emph {Relationship
  between Systems of Impenetrable Bosons and Fermions in One Dimension}},\
  }\href {\doibase 10.1063/1.1703687} {\bibfield  {journal} {\bibinfo
  {journal} {Journal of Mathematical Physics}\ }\textbf {\bibinfo {volume}
  {1}},\ \bibinfo {pages} {516} (\bibinfo {year} {1960})}\BibitemShut {NoStop}%
\bibitem [{\citenamefont {Paredes}\ \emph {et~al.}(2004)\citenamefont
  {Paredes}, \citenamefont {Widera}, \citenamefont {Murg}, \citenamefont
  {Mandel}, \citenamefont {F{\"o}lling}, \citenamefont {Cirac}, \citenamefont
  {Shlyapnikov}, \citenamefont {H{\"a}nsch},\ and\ \citenamefont
  {Bloch}}]{Paredes2004}%
  \BibitemOpen
  \bibfield  {author} {\bibinfo {author} {\bibfnamefont {B.}~\bibnamefont
  {Paredes}}, \bibinfo {author} {\bibfnamefont {A.}~\bibnamefont {Widera}},
  \bibinfo {author} {\bibfnamefont {V.}~\bibnamefont {Murg}}, \bibinfo {author}
  {\bibfnamefont {O.}~\bibnamefont {Mandel}}, \bibinfo {author} {\bibfnamefont
  {S.}~\bibnamefont {F{\"o}lling}}, \bibinfo {author} {\bibfnamefont
  {I.}~\bibnamefont {Cirac}}, \bibinfo {author} {\bibfnamefont {G.~V.}\
  \bibnamefont {Shlyapnikov}}, \bibinfo {author} {\bibfnamefont {T.~W.}\
  \bibnamefont {H{\"a}nsch}}, \ and\ \bibinfo {author} {\bibfnamefont
  {I.}~\bibnamefont {Bloch}},\ }\bibfield  {title} {\bibinfo {title} {\emph
  {Tonks--Girardeau gas of ultracold atoms in an optical lattice}},\ }\href
  {https://doi.org/10.1038/nature02530} {\bibfield  {journal} {\bibinfo
  {journal} {Nature}\ }\textbf {\bibinfo {volume} {429}},\ \bibinfo {pages}
  {277 EP } (\bibinfo {year} {2004})}\BibitemShut {NoStop}%
\bibitem [{\citenamefont {Kinoshita}\ \emph {et~al.}(2004)\citenamefont
  {Kinoshita}, \citenamefont {Wenger},\ and\ \citenamefont
  {Weiss}}]{Kinoshita2004}%
  \BibitemOpen
  \bibfield  {author} {\bibinfo {author} {\bibfnamefont {T.}~\bibnamefont
  {Kinoshita}}, \bibinfo {author} {\bibfnamefont {T.}~\bibnamefont {Wenger}}, \
  and\ \bibinfo {author} {\bibfnamefont {D.~S.}\ \bibnamefont {Weiss}},\
  }\bibfield  {title} {\bibinfo {title} {\emph {Observation of a
  One-Dimensional Tonks-Girardeau Gas}},\ }\href {\doibase
  10.1126/science.1100700} {\bibfield  {journal} {\bibinfo  {journal}
  {Science}\ }\textbf {\bibinfo {volume} {305}},\ \bibinfo {pages} {1125}
  (\bibinfo {year} {2004})}\BibitemShut {NoStop}%
\bibitem [{\citenamefont {Laburthe~Tolra}\ \emph {et~al.}(2004)\citenamefont
  {Laburthe~Tolra}, \citenamefont {O'Hara}, \citenamefont {Huckans},
  \citenamefont {Phillips}, \citenamefont {Rolston},\ and\ \citenamefont
  {Porto}}]{Tolra2004}%
  \BibitemOpen
  \bibfield  {author} {\bibinfo {author} {\bibfnamefont {B.}~\bibnamefont
  {Laburthe~Tolra}}, \bibinfo {author} {\bibfnamefont {K.~M.}\ \bibnamefont
  {O'Hara}}, \bibinfo {author} {\bibfnamefont {J.~H.}\ \bibnamefont {Huckans}},
  \bibinfo {author} {\bibfnamefont {W.~D.}\ \bibnamefont {Phillips}}, \bibinfo
  {author} {\bibfnamefont {S.~L.}\ \bibnamefont {Rolston}}, \ and\ \bibinfo
  {author} {\bibfnamefont {J.~V.}\ \bibnamefont {Porto}},\ }\bibfield  {title}
  {\bibinfo {title} {\emph {Observation of Reduced Three-Body Recombination in
  a Correlated 1D Degenerate Bose Gas}},\ }\href {\doibase
  10.1103/PhysRevLett.92.190401} {\bibfield  {journal} {\bibinfo  {journal}
  {Phys. Rev. Lett.}\ }\textbf {\bibinfo {volume} {92}},\ \bibinfo {pages}
  {190401} (\bibinfo {year} {2004})}\BibitemShut {NoStop}%
\bibitem [{\citenamefont {Kinoshita}\ \emph {et~al.}(2005)\citenamefont
  {Kinoshita}, \citenamefont {Wenger},\ and\ \citenamefont
  {Weiss}}]{Kinoshita2005}%
  \BibitemOpen
  \bibfield  {author} {\bibinfo {author} {\bibfnamefont {T.}~\bibnamefont
  {Kinoshita}}, \bibinfo {author} {\bibfnamefont {T.}~\bibnamefont {Wenger}}, \
  and\ \bibinfo {author} {\bibfnamefont {D.~S.}\ \bibnamefont {Weiss}},\
  }\bibfield  {title} {\bibinfo {title} {\emph {Local Pair Correlations in
  One-Dimensional Bose Gases}},\ }\href {\doibase
  10.1103/PhysRevLett.95.190406} {\bibfield  {journal} {\bibinfo  {journal}
  {Phys. Rev. Lett.}\ }\textbf {\bibinfo {volume} {95}},\ \bibinfo {pages}
  {190406} (\bibinfo {year} {2005})}\BibitemShut {NoStop}%
\bibitem [{\citenamefont {Haller}\ \emph {et~al.}(2011)\citenamefont {Haller},
  \citenamefont {Rabie}, \citenamefont {Mark}, \citenamefont {Danzl},
  \citenamefont {Hart}, \citenamefont {Lauber}, \citenamefont {Pupillo},\ and\
  \citenamefont {N\"agerl}}]{Haller2011}%
  \BibitemOpen
  \bibfield  {author} {\bibinfo {author} {\bibfnamefont {E.}~\bibnamefont
  {Haller}}, \bibinfo {author} {\bibfnamefont {M.}~\bibnamefont {Rabie}},
  \bibinfo {author} {\bibfnamefont {M.~J.}\ \bibnamefont {Mark}}, \bibinfo
  {author} {\bibfnamefont {J.~G.}\ \bibnamefont {Danzl}}, \bibinfo {author}
  {\bibfnamefont {R.}~\bibnamefont {Hart}}, \bibinfo {author} {\bibfnamefont
  {K.}~\bibnamefont {Lauber}}, \bibinfo {author} {\bibfnamefont
  {G.}~\bibnamefont {Pupillo}}, \ and\ \bibinfo {author} {\bibfnamefont
  {H.-C.}\ \bibnamefont {N\"agerl}},\ }\bibfield  {title} {\bibinfo {title}
  {\emph {Three-Body Correlation Functions and Recombination Rates for Bosons
  in Three Dimensions and One Dimension}},\ }\href {\doibase
  10.1103/PhysRevLett.107.230404} {\bibfield  {journal} {\bibinfo  {journal}
  {Phys. Rev. Lett.}\ }\textbf {\bibinfo {volume} {107}},\ \bibinfo {pages}
  {230404} (\bibinfo {year} {2011})}\BibitemShut {NoStop}%
\bibitem [{\citenamefont {Jacqmin}\ \emph {et~al.}(2011)\citenamefont
  {Jacqmin}, \citenamefont {Armijo}, \citenamefont {Berrada}, \citenamefont
  {Kheruntsyan},\ and\ \citenamefont {Bouchoule}}]{Jacqmin2011}%
  \BibitemOpen
  \bibfield  {author} {\bibinfo {author} {\bibfnamefont {T.}~\bibnamefont
  {Jacqmin}}, \bibinfo {author} {\bibfnamefont {J.}~\bibnamefont {Armijo}},
  \bibinfo {author} {\bibfnamefont {T.}~\bibnamefont {Berrada}}, \bibinfo
  {author} {\bibfnamefont {K.~V.}\ \bibnamefont {Kheruntsyan}}, \ and\ \bibinfo
  {author} {\bibfnamefont {I.}~\bibnamefont {Bouchoule}},\ }\bibfield  {title}
  {\bibinfo {title} {\emph {Sub-Poissonian Fluctuations in a 1D Bose Gas: From
  the Quantum Quasicondensate to the Strongly Interacting Regime}},\ }\href
  {\doibase 10.1103/PhysRevLett.106.230405} {\bibfield  {journal} {\bibinfo
  {journal} {Phys. Rev. Lett.}\ }\textbf {\bibinfo {volume} {106}},\ \bibinfo
  {pages} {230405} (\bibinfo {year} {2011})}\BibitemShut {NoStop}%
\bibitem [{\citenamefont {Guarrera}\ \emph {et~al.}(2012)\citenamefont
  {Guarrera}, \citenamefont {Muth}, \citenamefont {Labouvie}, \citenamefont
  {Vogler}, \citenamefont {Barontini}, \citenamefont {Fleischhauer},\ and\
  \citenamefont {Ott}}]{Guarrera2012}%
  \BibitemOpen
  \bibfield  {author} {\bibinfo {author} {\bibfnamefont {V.}~\bibnamefont
  {Guarrera}}, \bibinfo {author} {\bibfnamefont {D.}~\bibnamefont {Muth}},
  \bibinfo {author} {\bibfnamefont {R.}~\bibnamefont {Labouvie}}, \bibinfo
  {author} {\bibfnamefont {A.}~\bibnamefont {Vogler}}, \bibinfo {author}
  {\bibfnamefont {G.}~\bibnamefont {Barontini}}, \bibinfo {author}
  {\bibfnamefont {M.}~\bibnamefont {Fleischhauer}}, \ and\ \bibinfo {author}
  {\bibfnamefont {H.}~\bibnamefont {Ott}},\ }\bibfield  {title} {\bibinfo
  {title} {\emph {Spatiotemporal fermionization of strongly interacting
  one-dimensional bosons}},\ }\href {\doibase 10.1103/PhysRevA.86.021601}
  {\bibfield  {journal} {\bibinfo  {journal} {Phys. Rev. A}\ }\textbf {\bibinfo
  {volume} {86}},\ \bibinfo {pages} {021601(R)} (\bibinfo {year}
  {2012})}\BibitemShut {NoStop}%
\bibitem [{\citenamefont {Lieb}\ and\ \citenamefont
  {Liniger}(1963)}]{Lieb1963}%
  \BibitemOpen
  \bibfield  {author} {\bibinfo {author} {\bibfnamefont {E.~H.}\ \bibnamefont
  {Lieb}}\ and\ \bibinfo {author} {\bibfnamefont {W.}~\bibnamefont {Liniger}},\
  }\bibfield  {title} {\bibinfo {title} {\emph {Exact Analysis of an
  Interacting Bose Gas. I. The General Solution and the Ground State}},\ }\href
  {\doibase 10.1103/PhysRev.130.1605} {\bibfield  {journal} {\bibinfo
  {journal} {Phys. Rev.}\ }\textbf {\bibinfo {volume} {130}},\ \bibinfo {pages}
  {1605} (\bibinfo {year} {1963})}\BibitemShut {NoStop}%
\bibitem [{\citenamefont {Lieb}(1963)}]{Lieb1963II}%
  \BibitemOpen
  \bibfield  {author} {\bibinfo {author} {\bibfnamefont {E.~H.}\ \bibnamefont
  {Lieb}},\ }\bibfield  {title} {\bibinfo {title} {\emph {Exact Analysis of an
  Interacting Bose Gas. II. The Excitation Spectrum}},\ }\href {\doibase
  10.1103/PhysRev.130.1616} {\bibfield  {journal} {\bibinfo  {journal} {Phys.
  Rev.}\ }\textbf {\bibinfo {volume} {130}},\ \bibinfo {pages} {1616} (\bibinfo
  {year} {1963})}\BibitemShut {NoStop}%
\bibitem [{\citenamefont {De~Rosi}\ \emph {et~al.}(2017)\citenamefont
  {De~Rosi}, \citenamefont {Astrakharchik},\ and\ \citenamefont
  {Stringari}}]{DeRosi2017}%
  \BibitemOpen
  \bibfield  {author} {\bibinfo {author} {\bibfnamefont {G.}~\bibnamefont
  {De~Rosi}}, \bibinfo {author} {\bibfnamefont {G.~E.}\ \bibnamefont
  {Astrakharchik}}, \ and\ \bibinfo {author} {\bibfnamefont {S.}~\bibnamefont
  {Stringari}},\ }\bibfield  {title} {\bibinfo {title} {\emph {Thermodynamic
  behavior of a one-dimensional Bose gas at low temperature}},\ }\href
  {\doibase 10.1103/PhysRevA.96.013613} {\bibfield  {journal} {\bibinfo
  {journal} {Phys. Rev. A}\ }\textbf {\bibinfo {volume} {96}},\ \bibinfo
  {pages} {013613} (\bibinfo {year} {2017})}\BibitemShut {NoStop}%
\bibitem [{\citenamefont {Yang}\ and\ \citenamefont {Yang}(1969)}]{Yang1969}%
  \BibitemOpen
  \bibfield  {author} {\bibinfo {author} {\bibfnamefont {C.~N.}\ \bibnamefont
  {Yang}}\ and\ \bibinfo {author} {\bibfnamefont {C.~P.}\ \bibnamefont
  {Yang}},\ }\bibfield  {title} {\bibinfo {title} {\emph {Thermodynamics of a
  One-Dimensional System of Bosons with Repulsive Delta-Function
  Interaction}},\ }\href {\doibase 10.1063/1.1664947} {\bibfield  {journal}
  {\bibinfo  {journal} {Journal of Mathematical Physics}\ }\textbf {\bibinfo
  {volume} {10}},\ \bibinfo {pages} {1115} (\bibinfo {year}
  {1969})}\BibitemShut {NoStop}%
\bibitem [{\citenamefont {Yang}(1970)}]{Yang1970}%
  \BibitemOpen
  \bibfield  {author} {\bibinfo {author} {\bibfnamefont {C.~P.}\ \bibnamefont
  {Yang}},\ }\bibfield  {title} {\bibinfo {title} {\emph {One-Dimensional
  System of Bosons with Repulsive $\ensuremath{\delta}$-Function Interactions
  at a Finite Temperature $T$}},\ }\href {\doibase 10.1103/PhysRevA.2.154}
  {\bibfield  {journal} {\bibinfo  {journal} {Phys. Rev. A}\ }\textbf {\bibinfo
  {volume} {2}},\ \bibinfo {pages} {154} (\bibinfo {year} {1970})}\BibitemShut
  {NoStop}%
\bibitem [{\citenamefont {Braaten}\ \emph {et~al.}(2011)\citenamefont
  {Braaten}, \citenamefont {Kang},\ and\ \citenamefont
  {Platter}}]{Braaten2011}%
  \BibitemOpen
  \bibfield  {author} {\bibinfo {author} {\bibfnamefont {E.}~\bibnamefont
  {Braaten}}, \bibinfo {author} {\bibfnamefont {D.}~\bibnamefont {Kang}}, \
  and\ \bibinfo {author} {\bibfnamefont {L.}~\bibnamefont {Platter}},\
  }\bibfield  {title} {\bibinfo {title} {\emph {Universal Relations for
  Identical Bosons from Three-Body Physics}},\ }\href {\doibase
  10.1103/PhysRevLett.106.153005} {\bibfield  {journal} {\bibinfo  {journal}
  {Phys. Rev. Lett.}\ }\textbf {\bibinfo {volume} {106}},\ \bibinfo {pages}
  {153005} (\bibinfo {year} {2011})}\BibitemShut {NoStop}%
\bibitem [{\citenamefont {Yao}\ \emph {et~al.}(2018)\citenamefont {Yao},
  \citenamefont {Cl\'ement}, \citenamefont {Minguzzi}, \citenamefont
  {Vignolo},\ and\ \citenamefont {Sanchez-Palencia}}]{Yao2018}%
  \BibitemOpen
  \bibfield  {author} {\bibinfo {author} {\bibfnamefont {H.}~\bibnamefont
  {Yao}}, \bibinfo {author} {\bibfnamefont {D.}~\bibnamefont {Cl\'ement}},
  \bibinfo {author} {\bibfnamefont {A.}~\bibnamefont {Minguzzi}}, \bibinfo
  {author} {\bibfnamefont {P.}~\bibnamefont {Vignolo}}, \ and\ \bibinfo
  {author} {\bibfnamefont {L.}~\bibnamefont {Sanchez-Palencia}},\ }\bibfield
  {title} {\bibinfo {title} {\emph {Tan's Contact for Trapped Lieb-Liniger
  Bosons at Finite Temperature}},\ }\href {\doibase
  10.1103/PhysRevLett.121.220402} {\bibfield  {journal} {\bibinfo  {journal}
  {Phys. Rev. Lett.}\ }\textbf {\bibinfo {volume} {121}},\ \bibinfo {pages}
  {220402} (\bibinfo {year} {2018})}\BibitemShut {NoStop}%
\bibitem [{\citenamefont {De~Rosi}\ \emph {et~al.}(2019)\citenamefont
  {De~Rosi}, \citenamefont {Massignan}, \citenamefont {Lewenstein},\ and\
  \citenamefont {Astrakharchik}}]{DeRosi2019}%
  \BibitemOpen
  \bibfield  {author} {\bibinfo {author} {\bibfnamefont {G.}~\bibnamefont
  {De~Rosi}}, \bibinfo {author} {\bibfnamefont {P.}~\bibnamefont {Massignan}},
  \bibinfo {author} {\bibfnamefont {M.}~\bibnamefont {Lewenstein}}, \ and\
  \bibinfo {author} {\bibfnamefont {G.~E.}\ \bibnamefont {Astrakharchik}},\
  }\bibfield  {title} {\bibinfo {title} {\emph {Beyond-Luttinger-liquid
  thermodynamics of a one-dimensional Bose gas with repulsive contact
  interactions}},\ }\href {\doibase 10.1103/PhysRevResearch.1.033083}
  {\bibfield  {journal} {\bibinfo  {journal} {Phys. Rev. Research}\ }\textbf
  {\bibinfo {volume} {1}},\ \bibinfo {pages} {033083} (\bibinfo {year}
  {2019})}\BibitemShut {NoStop}%
\bibitem [{\citenamefont {De~Rosi}\ \emph {et~al.}(2023)\citenamefont
  {De~Rosi}, \citenamefont {Rota}, \citenamefont {Astrakharchik},\ and\
  \citenamefont {Boronat}}]{DeRosi2022}%
  \BibitemOpen
  \bibfield  {author} {\bibinfo {author} {\bibfnamefont {G.}~\bibnamefont
  {De~Rosi}}, \bibinfo {author} {\bibfnamefont {R.}~\bibnamefont {Rota}},
  \bibinfo {author} {\bibfnamefont {G.~E.}\ \bibnamefont {Astrakharchik}}, \
  and\ \bibinfo {author} {\bibfnamefont {J.}~\bibnamefont {Boronat}},\
  }\bibfield  {title} {\bibinfo {title} {\emph {Correlation properties of a
  one-dimensional repulsive Bose gas at finite temperature}},\ }\href {\doibase
  10.1088/1367-2630/acc6e6} {\bibfield  {journal} {\bibinfo  {journal} {New J.
  Phys.}\ }\textbf {\bibinfo {volume} {25}},\ \bibinfo {pages} {043002}
  (\bibinfo {year} {2023})}\BibitemShut {NoStop}%
\bibitem [{\citenamefont {Astrakharchik}\ \emph {et~al.}(2006)\citenamefont
  {Astrakharchik}, \citenamefont {Gangardt}, \citenamefont {Lozovik},\ and\
  \citenamefont {Sorokin}}]{Astrakharchik2006II}%
  \BibitemOpen
  \bibfield  {author} {\bibinfo {author} {\bibfnamefont {G.~E.}\ \bibnamefont
  {Astrakharchik}}, \bibinfo {author} {\bibfnamefont {D.~M.}\ \bibnamefont
  {Gangardt}}, \bibinfo {author} {\bibfnamefont {Y.~E.}\ \bibnamefont
  {Lozovik}}, \ and\ \bibinfo {author} {\bibfnamefont {I.~A.}\ \bibnamefont
  {Sorokin}},\ }\bibfield  {title} {\bibinfo {title} {\emph {Off-diagonal
  correlations of the Calogero-Sutherland model}},\ }\href {\doibase
  10.1103/PhysRevE.74.021105} {\bibfield  {journal} {\bibinfo  {journal} {Phys.
  Rev. E}\ }\textbf {\bibinfo {volume} {74}},\ \bibinfo {pages} {021105}
  (\bibinfo {year} {2006})}\BibitemShut {NoStop}%
\bibitem [{\citenamefont {Feynman}(1939)}]{Feynman1939}%
  \BibitemOpen
  \bibfield  {author} {\bibinfo {author} {\bibfnamefont {R.~P.}\ \bibnamefont
  {Feynman}},\ }\bibfield  {title} {\bibinfo {title} {\emph {Forces in
  Molecules}},\ }\href {\doibase 10.1103/PhysRev.56.340} {\bibfield  {journal}
  {\bibinfo  {journal} {Phys. Rev.}\ }\textbf {\bibinfo {volume} {56}},\
  \bibinfo {pages} {340} (\bibinfo {year} {1939})}\BibitemShut {NoStop}%
\bibitem [{\citenamefont {Yurovsky}\ \emph {et~al.}(2008)\citenamefont
  {Yurovsky}, \citenamefont {Olshanii},\ and\ \citenamefont
  {Weiss}}]{Yurovsky2008}%
  \BibitemOpen
  \bibfield  {author} {\bibinfo {author} {\bibfnamefont {V.~A.}\ \bibnamefont
  {Yurovsky}}, \bibinfo {author} {\bibfnamefont {M.}~\bibnamefont {Olshanii}},
  \ and\ \bibinfo {author} {\bibfnamefont {D.~S.}\ \bibnamefont {Weiss}},\
  }\bibfield  {title} {\bibinfo {title} {\emph {Collisions, correlations, and
  integrability in atom waveguides}}\ }(\bibinfo  {publisher} {Academic
  Press},\ \bibinfo {year} {2008})\ pp.\ \bibinfo {pages} {61--138}\BibitemShut
  {NoStop}%
\bibitem [{\citenamefont {Dunjko}\ and\ \citenamefont
  {Olshanii}(2011)}]{Dunjko2011}%
  \BibitemOpen
  \bibfield  {author} {\bibinfo {author} {\bibfnamefont {V.}~\bibnamefont
  {Dunjko}}\ and\ \bibinfo {author} {\bibfnamefont {M.}~\bibnamefont
  {Olshanii}},\ }\bibfield  {title} {\bibinfo {title} {\emph {A
  Hermite-Pad$\acute{e}$ perspective on the renormalization group, with an
  application to the correlation function of Lieb-Liniger gas}},\ }\href
  {\doibase 10.1088/1751-8113/44/5/055206} {\bibfield  {journal} {\bibinfo
  {journal} {Journal of Physics A: Mathematical and Theoretical}\ }\textbf
  {\bibinfo {volume} {44}},\ \bibinfo {pages} {055206} (\bibinfo {year}
  {2011})}\BibitemShut {NoStop}%
\bibitem [{\citenamefont {Olshanii}\ \emph {et~al.}(2017)\citenamefont
  {Olshanii}, \citenamefont {Dunjko}, \citenamefont {Minguzzi},\ and\
  \citenamefont {Lang}}]{Olshanii2017}%
  \BibitemOpen
  \bibfield  {author} {\bibinfo {author} {\bibfnamefont {M.}~\bibnamefont
  {Olshanii}}, \bibinfo {author} {\bibfnamefont {V.}~\bibnamefont {Dunjko}},
  \bibinfo {author} {\bibfnamefont {A.}~\bibnamefont {Minguzzi}}, \ and\
  \bibinfo {author} {\bibfnamefont {G.}~\bibnamefont {Lang}},\ }\bibfield
  {title} {\bibinfo {title} {\emph {Connection between nonlocal one-body and
  local three-body correlations of the Lieb-Liniger model}},\ }\href {\doibase
  10.1103/PhysRevA.96.033624} {\bibfield  {journal} {\bibinfo  {journal} {Phys.
  Rev. A}\ }\textbf {\bibinfo {volume} {96}},\ \bibinfo {pages} {033624}
  (\bibinfo {year} {2017})}\BibitemShut {NoStop}%
\bibitem [{\citenamefont {Cazalilla}(2004)}]{Cazalilla2004}%
  \BibitemOpen
  \bibfield  {author} {\bibinfo {author} {\bibfnamefont {M.~A.}\ \bibnamefont
  {Cazalilla}},\ }\bibfield  {title} {\bibinfo {title} {\emph {Bosonizing
  one-dimensional cold atomic gases}},\ }\href {\doibase
  10.1088/0953-4075/37/7/051} {\bibfield  {journal} {\bibinfo  {journal}
  {Journal of Physics B: Atomic, Molecular and Optical Physics}\ }\textbf
  {\bibinfo {volume} {37}},\ \bibinfo {pages} {S1} (\bibinfo {year}
  {2004})}\BibitemShut {NoStop}%
\bibitem [{Note1()}]{Note1}%
  \BibitemOpen
  \bibinfo {note} {See Supplemental Material at [...] for additional results of
  the one-body density matrix and the momentum distribution, including the
  Maxwell-Boltzmann regime of the classical ideal gas at high temperatures. The
  Supplemental Material includes Refs.~\cite {Cazalilla2004, Esteve2006,
  DeRosi2022}.}\BibitemShut {Stop}%
\bibitem [{\citenamefont {Astrakharchik}\ and\ \citenamefont
  {Giorgini}(2003)}]{Astrakharchik2003}%
  \BibitemOpen
  \bibfield  {author} {\bibinfo {author} {\bibfnamefont {G.~E.}\ \bibnamefont
  {Astrakharchik}}\ and\ \bibinfo {author} {\bibfnamefont {S.}~\bibnamefont
  {Giorgini}},\ }\bibfield  {title} {\bibinfo {title} {\emph {Correlation
  functions and momentum distribution of one-dimensional Bose systems}},\
  }\href {\doibase 10.1103/PhysRevA.68.031602} {\bibfield  {journal} {\bibinfo
  {journal} {Phys. Rev. A}\ }\textbf {\bibinfo {volume} {68}},\ \bibinfo
  {pages} {031602(R)} (\bibinfo {year} {2003})}\BibitemShut {NoStop}%
\bibitem [{\citenamefont {Astrakharchik}\ and\ \citenamefont
  {Giorgini}(2006)}]{Astrakharchik2006}%
  \BibitemOpen
  \bibfield  {author} {\bibinfo {author} {\bibfnamefont {G.~E.}\ \bibnamefont
  {Astrakharchik}}\ and\ \bibinfo {author} {\bibfnamefont {S.}~\bibnamefont
  {Giorgini}},\ }\bibfield  {title} {\bibinfo {title} {\emph {Correlation
  functions of a Lieb{\textendash}Liniger Bose gas}},\ }\href {\doibase
  10.1088/0953-4075/39/10/s01} {\bibfield  {journal} {\bibinfo  {journal}
  {Journal of Physics B: Atomic, Molecular and Optical Physics}\ }\textbf
  {\bibinfo {volume} {39}},\ \bibinfo {pages} {S1} (\bibinfo {year}
  {2006})}\BibitemShut {NoStop}%
\bibitem [{\citenamefont {Drummond}\ \emph {et~al.}(2004)\citenamefont
  {Drummond}, \citenamefont {Deuar},\ and\ \citenamefont
  {Kheruntsyan}}]{Drummond2004}%
  \BibitemOpen
  \bibfield  {author} {\bibinfo {author} {\bibfnamefont {P.~D.}\ \bibnamefont
  {Drummond}}, \bibinfo {author} {\bibfnamefont {P.}~\bibnamefont {Deuar}}, \
  and\ \bibinfo {author} {\bibfnamefont {K.~V.}\ \bibnamefont {Kheruntsyan}},\
  }\bibfield  {title} {\bibinfo {title} {\emph {Canonical Bose Gas Simulations
  with Stochastic Gauges}},\ }\href {\doibase 10.1103/PhysRevLett.92.040405}
  {\bibfield  {journal} {\bibinfo  {journal} {Phys. Rev. Lett.}\ }\textbf
  {\bibinfo {volume} {92}},\ \bibinfo {pages} {040405} (\bibinfo {year}
  {2004})}\BibitemShut {NoStop}%
\bibitem [{\citenamefont {Mora}\ and\ \citenamefont {Castin}(2003)}]{Mora2003}%
  \BibitemOpen
  \bibfield  {author} {\bibinfo {author} {\bibfnamefont {C.}~\bibnamefont
  {Mora}}\ and\ \bibinfo {author} {\bibfnamefont {Y.}~\bibnamefont {Castin}},\
  }\bibfield  {title} {\bibinfo {title} {\emph {Extension of Bogoliubov theory
  to quasicondensates}},\ }\href {\doibase 10.1103/PhysRevA.67.053615}
  {\bibfield  {journal} {\bibinfo  {journal} {Phys. Rev. A}\ }\textbf {\bibinfo
  {volume} {67}},\ \bibinfo {pages} {053615} (\bibinfo {year}
  {2003})}\BibitemShut {NoStop}%
\bibitem [{\citenamefont {Cheng}\ \emph {et~al.}(2022)\citenamefont {Cheng},
  \citenamefont {Chen}, \citenamefont {Guan}, \citenamefont {Yang},\ and\
  \citenamefont {Lin}}]{Cheng2022}%
  \BibitemOpen
  \bibfield  {author} {\bibinfo {author} {\bibfnamefont {S.}~\bibnamefont
  {Cheng}}, \bibinfo {author} {\bibfnamefont {Y.-Y.}\ \bibnamefont {Chen}},
  \bibinfo {author} {\bibfnamefont {X.-W.}\ \bibnamefont {Guan}}, \bibinfo
  {author} {\bibfnamefont {W.-L.}\ \bibnamefont {Yang}}, \ and\ \bibinfo
  {author} {\bibfnamefont {H.-Q.}\ \bibnamefont {Lin}},\ }\bibfield  {title}
  {\bibinfo {title} {\emph {One-body dynamical correlation function of
  Lieb-Liniger model at finite temperature}},\ }\href
  {https://arxiv.org/abs/2211.00282} {\bibfield  {journal} {\bibinfo  {journal}
  {arXiv:2211.00282}\ } (\bibinfo {year} {2022})}\BibitemShut {NoStop}%
\bibitem [{\citenamefont {Rauer}\ \emph {et~al.}(2018)\citenamefont {Rauer},
  \citenamefont {Erne}, \citenamefont {Schweigler}, \citenamefont {Cataldini},
  \citenamefont {Tajik},\ and\ \citenamefont {Schmiedmayer}}]{Rauer2018}%
  \BibitemOpen
  \bibfield  {author} {\bibinfo {author} {\bibfnamefont {B.}~\bibnamefont
  {Rauer}}, \bibinfo {author} {\bibfnamefont {S.}~\bibnamefont {Erne}},
  \bibinfo {author} {\bibfnamefont {T.}~\bibnamefont {Schweigler}}, \bibinfo
  {author} {\bibfnamefont {F.}~\bibnamefont {Cataldini}}, \bibinfo {author}
  {\bibfnamefont {M.}~\bibnamefont {Tajik}}, \ and\ \bibinfo {author}
  {\bibfnamefont {J.}~\bibnamefont {Schmiedmayer}},\ }\bibfield  {title}
  {\bibinfo {title} {\emph {Recurrences in an isolated quantum many-body
  system}},\ }\href {\doibase 10.1126/science.aan7938} {\bibfield  {journal}
  {\bibinfo  {journal} {Science}\ }\textbf {\bibinfo {volume} {360}},\ \bibinfo
  {pages} {307} (\bibinfo {year} {2018})}\BibitemShut {NoStop}%
\bibitem [{\citenamefont {Chin}\ \emph {et~al.}(2010)\citenamefont {Chin},
  \citenamefont {Grimm}, \citenamefont {Julienne},\ and\ \citenamefont
  {Tiesinga}}]{Chin2010}%
  \BibitemOpen
  \bibfield  {author} {\bibinfo {author} {\bibfnamefont {C.}~\bibnamefont
  {Chin}}, \bibinfo {author} {\bibfnamefont {R.}~\bibnamefont {Grimm}},
  \bibinfo {author} {\bibfnamefont {P.}~\bibnamefont {Julienne}}, \ and\
  \bibinfo {author} {\bibfnamefont {E.}~\bibnamefont {Tiesinga}},\ }\bibfield
  {title} {\bibinfo {title} {\emph {Feshbach resonances in ultracold gases}},\
  }\href {\doibase 10.1103/RevModPhys.82.1225} {\bibfield  {journal} {\bibinfo
  {journal} {Rev. Mod. Phys.}\ }\textbf {\bibinfo {volume} {82}},\ \bibinfo
  {pages} {1225} (\bibinfo {year} {2010})}\BibitemShut {NoStop}%
\bibitem [{\citenamefont {M\o{}ller}\ \emph {et~al.}(2021)\citenamefont
  {M\o{}ller}, \citenamefont {Schweigler}, \citenamefont {Tajik}, \citenamefont
  {Sabino}, \citenamefont {Cataldini}, \citenamefont {Ji},\ and\ \citenamefont
  {Schmiedmayer}}]{Moller2021}%
  \BibitemOpen
  \bibfield  {author} {\bibinfo {author} {\bibfnamefont {F.}~\bibnamefont
  {M\o{}ller}}, \bibinfo {author} {\bibfnamefont {T.}~\bibnamefont
  {Schweigler}}, \bibinfo {author} {\bibfnamefont {M.}~\bibnamefont {Tajik}},
  \bibinfo {author} {\bibfnamefont {J.~a.}\ \bibnamefont {Sabino}}, \bibinfo
  {author} {\bibfnamefont {F.}~\bibnamefont {Cataldini}}, \bibinfo {author}
  {\bibfnamefont {S.-C.}\ \bibnamefont {Ji}}, \ and\ \bibinfo {author}
  {\bibfnamefont {J.}~\bibnamefont {Schmiedmayer}},\ }\bibfield  {title}
  {\bibinfo {title} {\emph {Thermometry of one-dimensional Bose gases with
  neural networks}},\ }\href {\doibase 10.1103/PhysRevA.104.043305} {\bibfield
  {journal} {\bibinfo  {journal} {Phys. Rev. A}\ }\textbf {\bibinfo {volume}
  {104}},\ \bibinfo {pages} {043305} (\bibinfo {year} {2021})}\BibitemShut
  {NoStop}%
\bibitem [{\citenamefont {Fabbri}\ \emph {et~al.}(2011)\citenamefont {Fabbri},
  \citenamefont {Cl\'ement}, \citenamefont {Fallani}, \citenamefont {Fort},\
  and\ \citenamefont {Inguscio}}]{Fabbri2011}%
  \BibitemOpen
  \bibfield  {author} {\bibinfo {author} {\bibfnamefont {N.}~\bibnamefont
  {Fabbri}}, \bibinfo {author} {\bibfnamefont {D.}~\bibnamefont {Cl\'ement}},
  \bibinfo {author} {\bibfnamefont {L.}~\bibnamefont {Fallani}}, \bibinfo
  {author} {\bibfnamefont {C.}~\bibnamefont {Fort}}, \ and\ \bibinfo {author}
  {\bibfnamefont {M.}~\bibnamefont {Inguscio}},\ }\bibfield  {title} {\bibinfo
  {title} {\emph {Momentum-resolved study of an array of one-dimensional
  strongly phase-fluctuating Bose gases}},\ }\href {\doibase
  10.1103/PhysRevA.83.031604} {\bibfield  {journal} {\bibinfo  {journal} {Phys.
  Rev. A}\ }\textbf {\bibinfo {volume} {83}},\ \bibinfo {pages} {031604(R)}
  (\bibinfo {year} {2011})}\BibitemShut {NoStop}%
\bibitem [{\citenamefont {Wilson}\ \emph {et~al.}(2020)\citenamefont {Wilson},
  \citenamefont {Malvania}, \citenamefont {Le}, \citenamefont {Zhang},
  \citenamefont {Rigol},\ and\ \citenamefont {Weiss}}]{Wilson2020}%
  \BibitemOpen
  \bibfield  {author} {\bibinfo {author} {\bibfnamefont {J.~M.}\ \bibnamefont
  {Wilson}}, \bibinfo {author} {\bibfnamefont {N.}~\bibnamefont {Malvania}},
  \bibinfo {author} {\bibfnamefont {Y.}~\bibnamefont {Le}}, \bibinfo {author}
  {\bibfnamefont {Y.}~\bibnamefont {Zhang}}, \bibinfo {author} {\bibfnamefont
  {M.}~\bibnamefont {Rigol}}, \ and\ \bibinfo {author} {\bibfnamefont {D.~S.}\
  \bibnamefont {Weiss}},\ }\bibfield  {title} {\bibinfo {title} {\emph
  {Observation of dynamical fermionization}},\ }\href {\doibase
  10.1126/science.aaz0242} {\bibfield  {journal} {\bibinfo  {journal}
  {Science}\ }\textbf {\bibinfo {volume} {367}},\ \bibinfo {pages} {1461}
  (\bibinfo {year} {2020})}\BibitemShut {NoStop}%
\bibitem [{\citenamefont {Malvania}\ \emph {et~al.}(2021)\citenamefont
  {Malvania}, \citenamefont {Zhang}, \citenamefont {Le}, \citenamefont
  {Dubail}, \citenamefont {Rigol},\ and\ \citenamefont {Weiss}}]{Malvania2021}%
  \BibitemOpen
  \bibfield  {author} {\bibinfo {author} {\bibfnamefont {N.}~\bibnamefont
  {Malvania}}, \bibinfo {author} {\bibfnamefont {Y.}~\bibnamefont {Zhang}},
  \bibinfo {author} {\bibfnamefont {Y.}~\bibnamefont {Le}}, \bibinfo {author}
  {\bibfnamefont {J.}~\bibnamefont {Dubail}}, \bibinfo {author} {\bibfnamefont
  {M.}~\bibnamefont {Rigol}}, \ and\ \bibinfo {author} {\bibfnamefont {D.~S.}\
  \bibnamefont {Weiss}},\ }\bibfield  {title} {\bibinfo {title} {\emph
  {Generalized hydrodynamics in strongly interacting 1D Bose gases}},\ }\href
  {\doibase 10.1126/science.abf0147} {\bibfield  {journal} {\bibinfo  {journal}
  {Science}\ }\textbf {\bibinfo {volume} {373}},\ \bibinfo {pages} {1129}
  (\bibinfo {year} {2021})}\BibitemShut {NoStop}%
\bibitem [{\citenamefont {Le}\ \emph {et~al.}(2023)\citenamefont {Le},
  \citenamefont {Zhang}, \citenamefont {Gopalakrishnan}, \citenamefont
  {Rigol},\ and\ \citenamefont {Weiss}}]{Le2023}%
  \BibitemOpen
  \bibfield  {author} {\bibinfo {author} {\bibfnamefont {Y.}~\bibnamefont
  {Le}}, \bibinfo {author} {\bibfnamefont {Y.}~\bibnamefont {Zhang}}, \bibinfo
  {author} {\bibfnamefont {S.}~\bibnamefont {Gopalakrishnan}}, \bibinfo
  {author} {\bibfnamefont {M.}~\bibnamefont {Rigol}}, \ and\ \bibinfo {author}
  {\bibfnamefont {D.~S.}\ \bibnamefont {Weiss}},\ }\bibfield  {title} {\bibinfo
  {title} {\emph {Observation of hydrodynamization and local prethermalization
  in 1D Bose gases}},\ }\href {\doibase 10.1038/s41586-023-05979-9} {\bibfield
  {journal} {\bibinfo  {journal} {Nature}\ }\textbf {\bibinfo {volume} {618}},\
  \bibinfo {pages} {494} (\bibinfo {year} {2023})}\BibitemShut {NoStop}%
\bibitem [{\citenamefont {Astrakharchik}\ \emph {et~al.}(2005)\citenamefont
  {Astrakharchik}, \citenamefont {Boronat}, \citenamefont {Casulleras},\ and\
  \citenamefont {Giorgini}}]{Astrakharchik2005}%
  \BibitemOpen
  \bibfield  {author} {\bibinfo {author} {\bibfnamefont {G.~E.}\ \bibnamefont
  {Astrakharchik}}, \bibinfo {author} {\bibfnamefont {J.}~\bibnamefont
  {Boronat}}, \bibinfo {author} {\bibfnamefont {J.}~\bibnamefont {Casulleras}},
  \ and\ \bibinfo {author} {\bibfnamefont {S.}~\bibnamefont {Giorgini}},\
  }\bibfield  {title} {\bibinfo {title} {\emph {Beyond the Tonks-Girardeau Gas:
  Strongly Correlated Regime in Quasi-One-Dimensional Bose Gases}},\ }\href
  {\doibase 10.1103/PhysRevLett.95.190407} {\bibfield  {journal} {\bibinfo
  {journal} {Phys. Rev. Lett.}\ }\textbf {\bibinfo {volume} {95}},\ \bibinfo
  {pages} {190407} (\bibinfo {year} {2005})}\BibitemShut {NoStop}%
\bibitem [{\citenamefont {Haller}\ \emph {et~al.}(2009)\citenamefont {Haller},
  \citenamefont {Gustavsson}, \citenamefont {Mark}, \citenamefont {Danzl},
  \citenamefont {Hart}, \citenamefont {Pupillo},\ and\ \citenamefont
  {N{\"a}gerl}}]{Haller2009}%
  \BibitemOpen
  \bibfield  {author} {\bibinfo {author} {\bibfnamefont {E.}~\bibnamefont
  {Haller}}, \bibinfo {author} {\bibfnamefont {M.}~\bibnamefont {Gustavsson}},
  \bibinfo {author} {\bibfnamefont {M.~J.}\ \bibnamefont {Mark}}, \bibinfo
  {author} {\bibfnamefont {J.~G.}\ \bibnamefont {Danzl}}, \bibinfo {author}
  {\bibfnamefont {R.}~\bibnamefont {Hart}}, \bibinfo {author} {\bibfnamefont
  {G.}~\bibnamefont {Pupillo}}, \ and\ \bibinfo {author} {\bibfnamefont
  {H.-C.}\ \bibnamefont {N{\"a}gerl}},\ }\bibfield  {title} {\bibinfo {title}
  {\emph {Realization of an Excited, Strongly Correlated Quantum Gas Phase}},\
  }\href {\doibase 10.1126/science.1175850} {\bibfield  {journal} {\bibinfo
  {journal} {Science}\ }\textbf {\bibinfo {volume} {325}},\ \bibinfo {pages}
  {1224} (\bibinfo {year} {2009})}\BibitemShut {NoStop}%
\bibitem [{\citenamefont {Fischer}\ \emph {et~al.}(2014)\citenamefont
  {Fischer}, \citenamefont {Savenko}, \citenamefont {Fraser}, \citenamefont
  {Holzinger}, \citenamefont {Brodbeck}, \citenamefont {Kamp}, \citenamefont
  {Shelykh}, \citenamefont {Schneider},\ and\ \citenamefont
  {H\"ofling}}]{Fischer2014}%
  \BibitemOpen
  \bibfield  {author} {\bibinfo {author} {\bibfnamefont {J.}~\bibnamefont
  {Fischer}}, \bibinfo {author} {\bibfnamefont {I.~G.}\ \bibnamefont
  {Savenko}}, \bibinfo {author} {\bibfnamefont {M.~D.}\ \bibnamefont {Fraser}},
  \bibinfo {author} {\bibfnamefont {S.}~\bibnamefont {Holzinger}}, \bibinfo
  {author} {\bibfnamefont {S.}~\bibnamefont {Brodbeck}}, \bibinfo {author}
  {\bibfnamefont {M.}~\bibnamefont {Kamp}}, \bibinfo {author} {\bibfnamefont
  {I.~A.}\ \bibnamefont {Shelykh}}, \bibinfo {author} {\bibfnamefont
  {C.}~\bibnamefont {Schneider}}, \ and\ \bibinfo {author} {\bibfnamefont
  {S.}~\bibnamefont {H\"ofling}},\ }\bibfield  {title} {\bibinfo {title} {\emph
  {Spatial Coherence Properties of One Dimensional Exciton-Polariton
  Condensates}},\ }\href {\doibase 10.1103/PhysRevLett.113.203902} {\bibfield
  {journal} {\bibinfo  {journal} {Phys. Rev. Lett.}\ }\textbf {\bibinfo
  {volume} {113}},\ \bibinfo {pages} {203902} (\bibinfo {year}
  {2014})}\BibitemShut {NoStop}%
\bibitem [{\citenamefont {Benhar}\ and\ \citenamefont
  {De~Rosi}(2017)}]{DeRosi2017II}%
  \BibitemOpen
  \bibfield  {author} {\bibinfo {author} {\bibfnamefont {O.}~\bibnamefont
  {Benhar}}\ and\ \bibinfo {author} {\bibfnamefont {G.}~\bibnamefont
  {De~Rosi}},\ }\bibfield  {title} {\bibinfo {title} {\emph {Superfluid Gap in
  Neutron Matter from a Microscopic Effective Interaction}},\ }\href {\doibase
  10.1007/s10909-017-1823-x} {\bibfield  {journal} {\bibinfo  {journal} {J Low
  Temp Phys}\ }\textbf {\bibinfo {volume} {189}},\ \bibinfo {pages} {250}
  (\bibinfo {year} {2017})}\BibitemShut {NoStop}%
\bibitem [{\citenamefont {De~Rosi}\ \emph {et~al.}(2021)\citenamefont
  {De~Rosi}, \citenamefont {Astrakharchik},\ and\ \citenamefont
  {Massignan}}]{DeRosi2021}%
  \BibitemOpen
  \bibfield  {author} {\bibinfo {author} {\bibfnamefont {G.}~\bibnamefont
  {De~Rosi}}, \bibinfo {author} {\bibfnamefont {G.~E.}\ \bibnamefont
  {Astrakharchik}}, \ and\ \bibinfo {author} {\bibfnamefont {P.}~\bibnamefont
  {Massignan}},\ }\bibfield  {title} {\bibinfo {title} {\emph {Thermal
  instability, evaporation, and thermodynamics of one-dimensional liquids in
  weakly interacting Bose-Bose mixtures}},\ }\href {\doibase
  10.1103/PhysRevA.103.043316} {\bibfield  {journal} {\bibinfo  {journal}
  {Phys. Rev. A}\ }\textbf {\bibinfo {volume} {103}},\ \bibinfo {pages}
  {043316} (\bibinfo {year} {2021})}\BibitemShut {NoStop}%
\bibitem [{\citenamefont {Hofmann}\ and\ \citenamefont
  {Zwerger}(2021)}]{Hofmann2021}%
  \BibitemOpen
  \bibfield  {author} {\bibinfo {author} {\bibfnamefont {J.}~\bibnamefont
  {Hofmann}}\ and\ \bibinfo {author} {\bibfnamefont {W.}~\bibnamefont
  {Zwerger}},\ }\bibfield  {title} {\bibinfo {title} {\emph {Universal
  relations for dipolar quantum gases}},\ }\href {\doibase
  10.1103/PhysRevResearch.3.013088} {\bibfield  {journal} {\bibinfo  {journal}
  {Phys. Rev. Res.}\ }\textbf {\bibinfo {volume} {3}},\ \bibinfo {pages}
  {013088} (\bibinfo {year} {2021})}\BibitemShut {NoStop}%
\bibitem [{\citenamefont {Li}\ \emph {et~al.}(2023)\citenamefont {Li},
  \citenamefont {Zhang}, \citenamefont {Yang}, \citenamefont {Lin},
  \citenamefont {Gopalakrishnan}, \citenamefont {Rigol},\ and\ \citenamefont
  {Lev}}]{Li2023}%
  \BibitemOpen
  \bibfield  {author} {\bibinfo {author} {\bibfnamefont {K.-Y.}\ \bibnamefont
  {Li}}, \bibinfo {author} {\bibfnamefont {Y.}~\bibnamefont {Zhang}}, \bibinfo
  {author} {\bibfnamefont {K.}~\bibnamefont {Yang}}, \bibinfo {author}
  {\bibfnamefont {K.-Y.}\ \bibnamefont {Lin}}, \bibinfo {author} {\bibfnamefont
  {S.}~\bibnamefont {Gopalakrishnan}}, \bibinfo {author} {\bibfnamefont
  {M.}~\bibnamefont {Rigol}}, \ and\ \bibinfo {author} {\bibfnamefont {B.~L.}\
  \bibnamefont {Lev}},\ }\bibfield  {title} {\bibinfo {title} {\emph {Rapidity
  and momentum distributions of one-dimensional dipolar quantum gases}},\
  }\href {\doibase 10.1103/PhysRevA.107.L061302} {\bibfield  {journal}
  {\bibinfo  {journal} {Phys. Rev. A}\ }\textbf {\bibinfo {volume} {107}},\
  \bibinfo {pages} {L061302} (\bibinfo {year} {2023})}\BibitemShut {NoStop}%
\bibitem [{\citenamefont {Del~Maestro}\ \emph {et~al.}(2022)\citenamefont
  {Del~Maestro}, \citenamefont {Nichols}, \citenamefont {Prisk}, \citenamefont
  {Warren},\ and\ \citenamefont {Sokol}}]{DelMaestro2022}%
  \BibitemOpen
  \bibfield  {author} {\bibinfo {author} {\bibfnamefont {A.}~\bibnamefont
  {Del~Maestro}}, \bibinfo {author} {\bibfnamefont {N.}~\bibnamefont
  {Nichols}}, \bibinfo {author} {\bibfnamefont {T.}~\bibnamefont {Prisk}},
  \bibinfo {author} {\bibfnamefont {G.}~\bibnamefont {Warren}}, \ and\ \bibinfo
  {author} {\bibfnamefont {P.~E.}\ \bibnamefont {Sokol}},\ }\bibfield  {title}
  {\bibinfo {title} {\emph {Experimental realization of one dimensional
  helium}},\ }\href {\doibase https://doi.org/10.1038/s41467-022-30752-3}
  {\bibfield  {journal} {\bibinfo  {journal} {Nature Comm.}\ }\textbf {\bibinfo
  {volume} {13}},\ \bibinfo {pages} {3168} (\bibinfo {year}
  {2022})}\BibitemShut {NoStop}%
\bibitem [{\citenamefont {Murthy}\ \emph {et~al.}(2015)\citenamefont {Murthy},
  \citenamefont {Boettcher}, \citenamefont {Bayha}, \citenamefont {Holzmann},
  \citenamefont {Kedar}, \citenamefont {Neidig}, \citenamefont {Ries},
  \citenamefont {Wenz}, \citenamefont {Z\"urn},\ and\ \citenamefont
  {Jochim}}]{Murthy2015}%
  \BibitemOpen
  \bibfield  {author} {\bibinfo {author} {\bibfnamefont {P.~A.}\ \bibnamefont
  {Murthy}}, \bibinfo {author} {\bibfnamefont {I.}~\bibnamefont {Boettcher}},
  \bibinfo {author} {\bibfnamefont {L.}~\bibnamefont {Bayha}}, \bibinfo
  {author} {\bibfnamefont {M.}~\bibnamefont {Holzmann}}, \bibinfo {author}
  {\bibfnamefont {D.}~\bibnamefont {Kedar}}, \bibinfo {author} {\bibfnamefont
  {M.}~\bibnamefont {Neidig}}, \bibinfo {author} {\bibfnamefont {M.~G.}\
  \bibnamefont {Ries}}, \bibinfo {author} {\bibfnamefont {A.~N.}\ \bibnamefont
  {Wenz}}, \bibinfo {author} {\bibfnamefont {G.}~\bibnamefont {Z\"urn}}, \ and\
  \bibinfo {author} {\bibfnamefont {S.}~\bibnamefont {Jochim}},\ }\bibfield
  {title} {\bibinfo {title} {\emph {Observation of the
  Berezinskii-Kosterlitz-Thouless Phase Transition in an Ultracold Fermi
  Gas}},\ }\href {\doibase 10.1103/PhysRevLett.115.010401} {\bibfield
  {journal} {\bibinfo  {journal} {Phys. Rev. Lett.}\ }\textbf {\bibinfo
  {volume} {115}},\ \bibinfo {pages} {010401} (\bibinfo {year}
  {2015})}\BibitemShut {NoStop}%
\bibitem [{\citenamefont {Esteve}\ \emph {et~al.}(2006)\citenamefont {Esteve},
  \citenamefont {Trebbia}, \citenamefont {Schumm}, \citenamefont {Aspect},
  \citenamefont {Westbrook},\ and\ \citenamefont {Bouchoule}}]{Esteve2006}%
  \BibitemOpen
  \bibfield  {author} {\bibinfo {author} {\bibfnamefont {J.}~\bibnamefont
  {Esteve}}, \bibinfo {author} {\bibfnamefont {J.-B.}\ \bibnamefont {Trebbia}},
  \bibinfo {author} {\bibfnamefont {T.}~\bibnamefont {Schumm}}, \bibinfo
  {author} {\bibfnamefont {A.}~\bibnamefont {Aspect}}, \bibinfo {author}
  {\bibfnamefont {C.~I.}\ \bibnamefont {Westbrook}}, \ and\ \bibinfo {author}
  {\bibfnamefont {I.}~\bibnamefont {Bouchoule}},\ }\bibfield  {title} {\bibinfo
  {title} {\emph {Observations of Density Fluctuations in an Elongated Bose
  Gas: Ideal Gas and Quasicondensate Regimes}},\ }\href {\doibase
  10.1103/PhysRevLett.96.130403} {\bibfield  {journal} {\bibinfo  {journal}
  {Phys. Rev. Lett.}\ }\textbf {\bibinfo {volume} {96}},\ \bibinfo {pages}
  {130403} (\bibinfo {year} {2006})}\BibitemShut {NoStop}%
\end{thebibliography}%


\begin{thebibliography}{3}%
\makeatletter
\providecommand \@ifxundefined [1]{%
 \@ifx{#1\undefined}
}%
\providecommand \@ifnum [1]{%
 \ifnum #1\expandafter \@firstoftwo
 \else \expandafter \@secondoftwo
 \fi
}%
\providecommand \@ifx [1]{%
 \ifx #1\expandafter \@firstoftwo
 \else \expandafter \@secondoftwo
 \fi
}%
\providecommand \natexlab [1]{#1}%
\providecommand \enquote  [1]{``#1''}%
\providecommand \bibnamefont  [1]{#1}%
\providecommand \bibfnamefont [1]{#1}%
\providecommand \citenamefont [1]{#1}%
\providecommand \href@noop [0]{\@secondoftwo}%
\providecommand \href [0]{\begingroup \@sanitize@url \@href}%
\providecommand \@href[1]{\@@startlink{#1}\@@href}%
\providecommand \@@href[1]{\endgroup#1\@@endlink}%
\providecommand \@sanitize@url [0]{\catcode `\\12\catcode `\$12\catcode
  `\&12\catcode `\#12\catcode `\^12\catcode `\_12\catcode `\%12\relax}%
\providecommand \@@startlink[1]{}%
\providecommand \@@endlink[0]{}%
\providecommand \url  [0]{\begingroup\@sanitize@url \@url }%
\providecommand \@url [1]{\endgroup\@href {#1}{\urlprefix }}%
\providecommand \urlprefix  [0]{URL }%
\providecommand \Eprint [0]{\href }%
\providecommand \doibase [0]{http://dx.doi.org/}%
\providecommand \selectlanguage [0]{\@gobble}%
\providecommand \bibinfo  [0]{\@secondoftwo}%
\providecommand \bibfield  [0]{\@secondoftwo}%
\providecommand \translation [1]{[#1]}%
\providecommand \BibitemOpen [0]{}%
\providecommand \bibitemStop [0]{}%
\providecommand \bibitemNoStop [0]{.\EOS\space}%
\providecommand \EOS [0]{\spacefactor3000\relax}%
\providecommand \BibitemShut  [1]{\csname bibitem#1\endcsname}%
\let\auto@bib@innerbib\@empty
\bibitem [{\citenamefont {De~Rosi}\ \emph {et~al.}(2023)\citenamefont
  {De~Rosi}, \citenamefont {Rota}, \citenamefont {Astrakharchik},\ and\
  \citenamefont {Boronat}}]{DeRosi2022}%
  \BibitemOpen
  \bibfield  {author} {\bibinfo {author} {\bibfnamefont {G.}~\bibnamefont
  {De~Rosi}}, \bibinfo {author} {\bibfnamefont {R.}~\bibnamefont {Rota}},
  \bibinfo {author} {\bibfnamefont {G.~E.}\ \bibnamefont {Astrakharchik}}, \
  and\ \bibinfo {author} {\bibfnamefont {J.}~\bibnamefont {Boronat}},\
  }\bibfield  {title} {\bibinfo {title} {\emph {Correlation properties of a
  one-dimensional repulsive Bose gas at finite temperature}},\ }\href {\doibase
  10.1088/1367-2630/acc6e6} {\bibfield  {journal} {\bibinfo  {journal} {New J.
  Phys.}\ }\textbf {\bibinfo {volume} {25}},\ \bibinfo {pages} {043002}
  (\bibinfo {year} {2023})}\BibitemShut {NoStop}%
\bibitem [{\citenamefont {Cazalilla}(2004)}]{Cazalilla2004}%
  \BibitemOpen
  \bibfield  {author} {\bibinfo {author} {\bibfnamefont {M.~A.}\ \bibnamefont
  {Cazalilla}},\ }\bibfield  {title} {\bibinfo {title} {\emph {Bosonizing
  one-dimensional cold atomic gases}},\ }\href {\doibase
  10.1088/0953-4075/37/7/051} {\bibfield  {journal} {\bibinfo  {journal}
  {Journal of Physics B: Atomic, Molecular and Optical Physics}\ }\textbf
  {\bibinfo {volume} {37}},\ \bibinfo {pages} {S1} (\bibinfo {year}
  {2004})}\BibitemShut {NoStop}%
\bibitem [{\citenamefont {Esteve}\ \emph {et~al.}(2006)\citenamefont {Esteve},
  \citenamefont {Trebbia}, \citenamefont {Schumm}, \citenamefont {Aspect},
  \citenamefont {Westbrook},\ and\ \citenamefont {Bouchoule}}]{Esteve2006}%
  \BibitemOpen
  \bibfield  {author} {\bibinfo {author} {\bibfnamefont {J.}~\bibnamefont
  {Esteve}}, \bibinfo {author} {\bibfnamefont {J.-B.}\ \bibnamefont {Trebbia}},
  \bibinfo {author} {\bibfnamefont {T.}~\bibnamefont {Schumm}}, \bibinfo
  {author} {\bibfnamefont {A.}~\bibnamefont {Aspect}}, \bibinfo {author}
  {\bibfnamefont {C.~I.}\ \bibnamefont {Westbrook}}, \ and\ \bibinfo {author}
  {\bibfnamefont {I.}~\bibnamefont {Bouchoule}},\ }\bibfield  {title} {\bibinfo
  {title} {\emph {Observations of Density Fluctuations in an Elongated Bose
  Gas: Ideal Gas and Quasicondensate Regimes}},\ }\href {\doibase
  10.1103/PhysRevLett.96.130403} {\bibfield  {journal} {\bibinfo  {journal}
  {Phys. Rev. Lett.}\ }\textbf {\bibinfo {volume} {96}},\ \bibinfo {pages}
  {130403} (\bibinfo {year} {2006})}\BibitemShut {NoStop}%
\end{thebibliography}%

 \renewcommand{\theequation}{S\arabic{equation}}
 \setcounter{equation}{0}
 \renewcommand{\thefigure}{S\arabic{figure}}
 \setcounter{figure}{0}
 \renewcommand{\thesection}{S\arabic{section}}
 \setcounter{section}{0}
 \onecolumngrid

\end{document}


\title{Supplemental Material for ``Thermal fading of the $1/k^4$-tail of the momentum distribution induced by the hole anomaly''
}

\author{Giulia De Rosi}
\email{giulia.de.rosi@upc.edu}
\affiliation{Departament de F\'isica, Universitat Polit\`ecnica de Catalunya, Campus Nord B4-B5, 08034 Barcelona, Spain}

\author{Grigori E. Astrakharchik}
\email{grigori.astrakharchik@upc.edu}
\affiliation{Departament de F\'isica, Universitat Polit\`ecnica de Catalunya, Campus Nord B4-B5, 08034 Barcelona, Spain}
\affiliation{Departament de F{\'i}sica Qu{\`a}ntica i Astrof{\'i}sica, Facultat de F{\'i}sica, Universitat de Barcelona, E-08028 Barcelona, Spain}

\author{Maxim Olshanii}
\affiliation{Department of Physics, University of Massachusetts Boston, Boston Massachusetts 02125, USA}

\author{Jordi Boronat}
\email{jordi.boronat@upc.edu}
\affiliation{Departament de F\'isica, Universitat Polit\`ecnica de Catalunya, Campus Nord B4-B5, 08034 Barcelona, Spain}

\date{\today}

\maketitle


 \renewcommand{\theequation}{S\arabic{equation}}
 \setcounter{equation}{0}
 \renewcommand{\thefigure}{S\arabic{figure}}
 \setcounter{figure}{0}
 \renewcommand{\thesection}{S\arabic{section}}
 \setcounter{section}{0}
 \onecolumngrid



In this Supplemental Material, we show with symbols the additional exact Path Integral Monte Carlo (PIMC) results for the one-body density matrix (OBDM) (Sec.~\ref{Sec:OBDM}) and the momentum distribution (Sec.~\ref{Sec:n(k)}) in a one-dimensional Bose gas with contact repulsive interaction.~We consider different characteristic values of the interaction strength $\gamma$ and temperature crossing the hole-anomaly threshold $T_A$ (see Fig.~1 in the main text) rescaled in units of the quantum degeneracy value $T_d = \hbar^2 n^2/\left(2 m k_B\right)$.~Our findings also include the Maxwell-Boltzmann (MB) regime describing the classical ideal gas at very high temperatures. 

\section{One-Body Density Matrix}
\label{Sec:OBDM}

\begin{figure}[h!]
\includegraphics[width=0.76\textwidth]{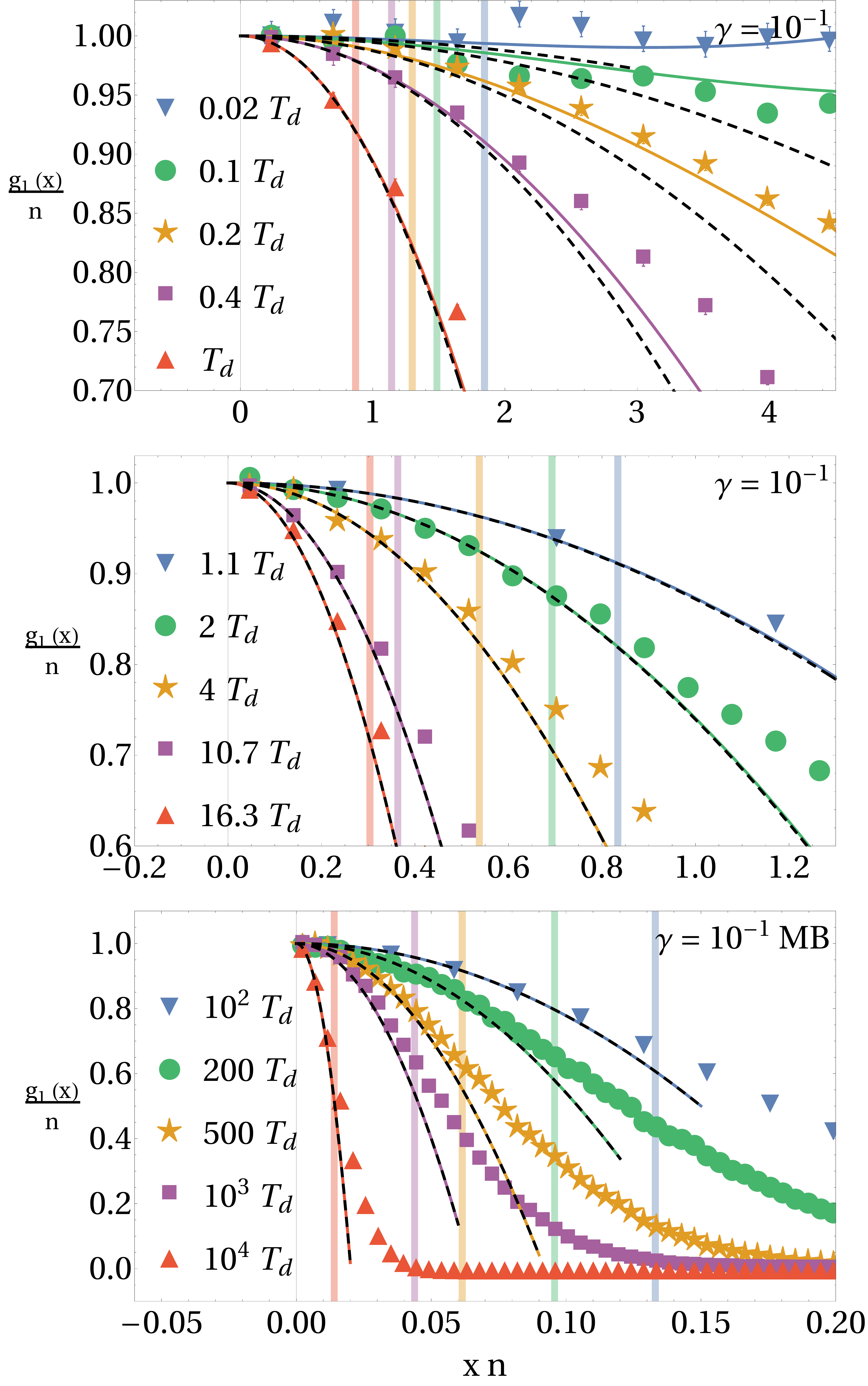}
\caption{One-body density matrix $g_1\left(x\right)/n$ as a function of the interparticle distance $x n$ ($n$ is the density) for weak interactions $\gamma = 10^{-1}$.~Different values of temperature in units of the quantum degeneracy value $T_d = \hbar^2 n^2/\left(2 m k_B\right)$ are reported.~Symbols denote exact Path Integral Monte Carlo results and their sizes are larger than the statistical error bars in some cases.~Solid lines represent the complete expansion at short distances (Eq.~(4) of the main text) calculated with exact thermal Bethe-Ansatz method.~Dashed black lines correspond to the same, but without the non-analytic cubic term (i.e. with $b_3 = 0$).~All curves are reported in increasing order of the temperature from small (top) to large (bottom) values in each panel.~The maximum distance for the validity of the short-range expansion $x_{\rm max} n = \left(\xi^{-1} + \sigma^{-1}\right)^{-1} n$ is represented by vertical lines from low (right) to high (left) temperature in each panel.~The lowest panel corresponds to the Maxwell-Boltzmann regime of the classical ideal gas at very high temperatures. 
}
\label{fig:OBDM01}
\end{figure}

\begin{figure}[h!]
\includegraphics[width=1\textwidth]{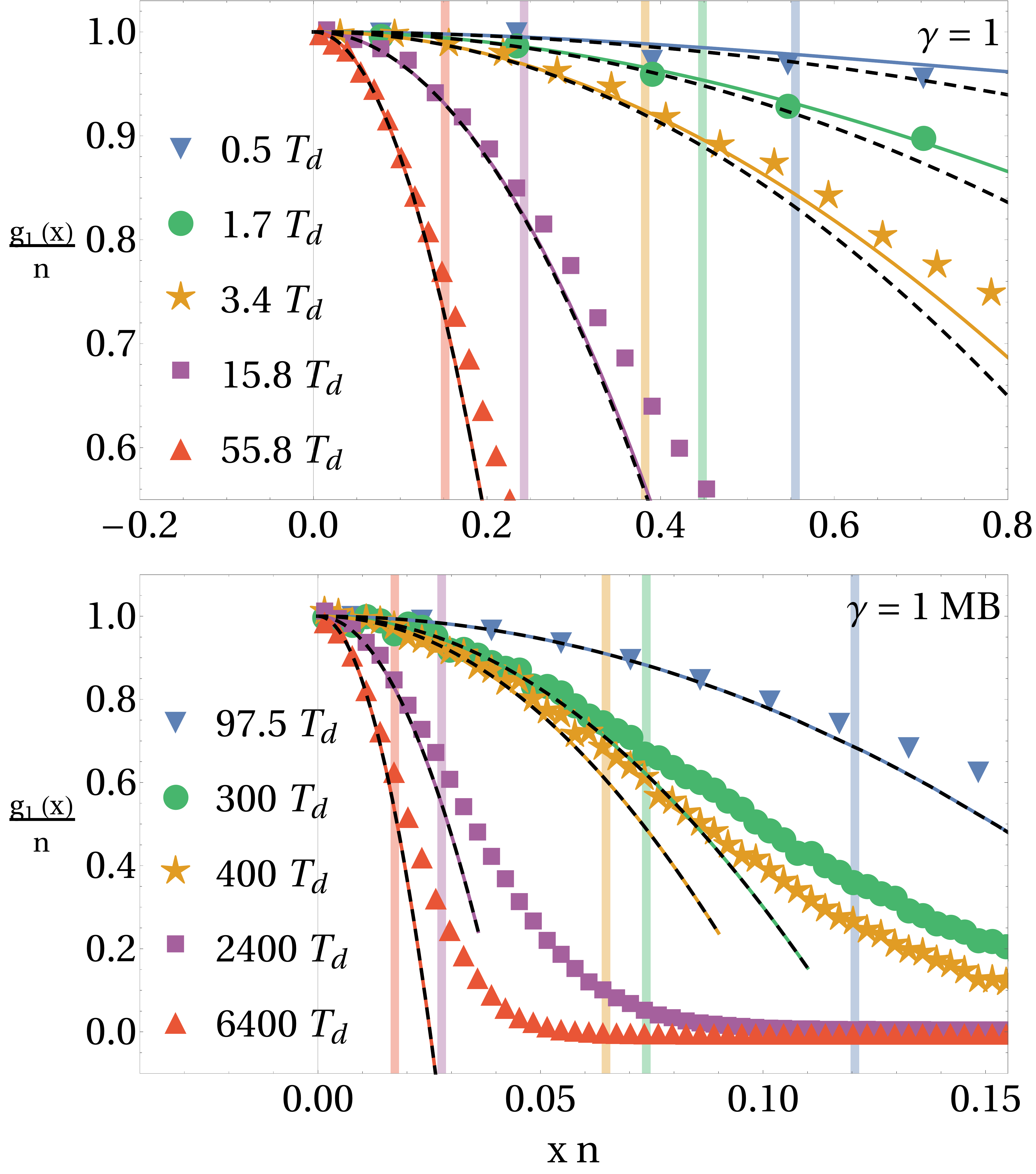}
\caption{OBDM for the intermediate interaction strength $\gamma = 1$, similarly as in Fig.~\ref{fig:OBDM01}.~Symbols denote PIMC results.~Solid lines represent the complete short-distance expansion (Eq.~(4) of the main text) calculated with TBA.~Dashed black lines correspond to the same expansion, but without the non-analytic cubic term (i.e.~with $b_3 = 0$).~All curves are reported for increasing temperature from small (top) to large (bottom) values in each panel.~The maximum distance $x_{\rm max} n$ is represented by vertical lines from low (right) to high (left) temperature in each panel.~The lower panel refers to the MB regime of the classical gas limit at very high temperatures.    
}
\label{fig:OBDM1}
\end{figure}

\begin{figure}[h!]
\includegraphics[width=0.76\textwidth]{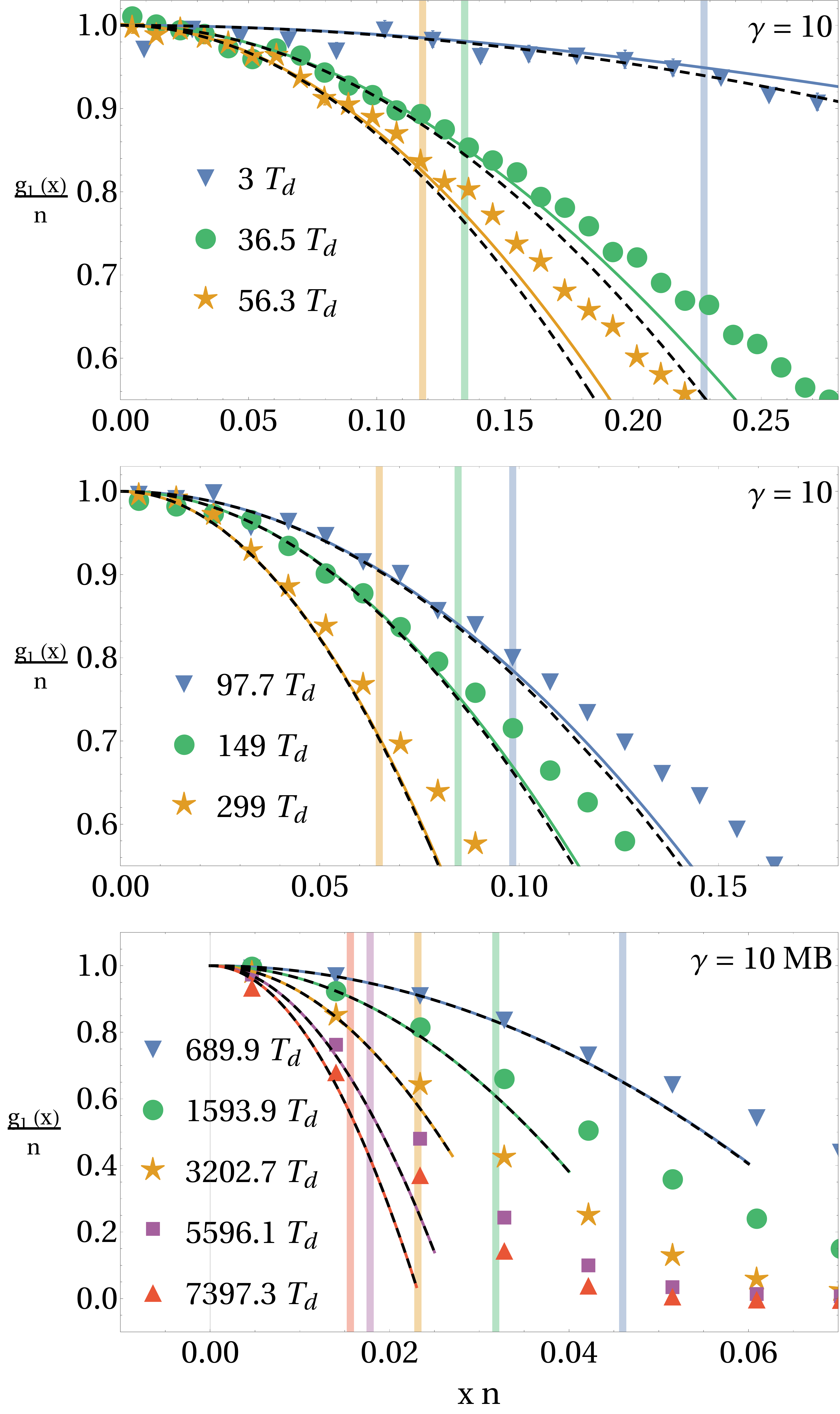}
\caption{OBDM for strong interactions $\gamma = 10$, similarly as in Fig.~\ref{fig:OBDM01}.~Symbols correspond to PIMC results.~Solid lines represent the complete expansion at short distances (Eq.~(4) of the main text) calculated with TBA.~Dashed black lines correspond to the same, but without the non-analytic cubic term (i.e.~with $b_3 = 0$).~All curves are reported in increasing order of the temperature from small (top) to large (bottom) values in each panel.~The maximum distance $x_{\rm max} n$ is represented by vertical lines from low (right) to high (left) temperature in each panel.~The lowest panel refers to the MB classical gas regime at very high temperatures. 
}
\label{fig:OBDM10}
\end{figure}

In Figs.~\ref{fig:OBDM01}-\ref{fig:OBDM10}, we report the one-body density matrix $g_1\left(x\right)$.~We compare the complete short-distance expansion of the OBDM (Eq.~(4) of the main text), which is denoted by colored solid lines, with the same result but without the non-analytic cubic term (whose coefficient is $b_3$) and corresponding to black dashed lines.~Both curves are calculated with exact thermal Bethe-Ansatz (TBA) method.~The maximal distance $x_{\rm max} = \left(\xi^{-1} + \sigma^{-1}\right)^{-1}$ (Eq.~(7) in the main text) for the validity of the expansion at short range is represented with vertical lines.~In the lowest panel of each figure, PIMC results have been previously checked to agree with the analytical Maxwell-Boltzmann classical gas limit $g_1\left(x \right)_{\rm MB}  = n e^{-x^2/(2\sigma^2)}$ \cite{DeRosi2022}.

In Fig.~\ref{fig:OBDM01}, we present the case of weak interactions with $\gamma = 10^{-1}$.~Each panel corresponds to different physical thermal regimes in order of increasing temperature values: hole-anomaly (upper panel), intermediate (middle) and MB (lower) regime. 

Fig.~\ref{fig:OBDM1} reports the intermediate value of the interaction strength $\gamma = 1$ in the anomaly (upper panel) and MB (lower) regime.  

Fig.~\ref{fig:OBDM10} shows $\gamma = 10$ for strong interactions, in the anomaly (upper panel), virial hard-core (middle) and MB (lower) regime.

Our approximation for $x_{\rm max}$ (Eq.~(7) in the main text) is valid along the whole interaction and temperature crossover of the system, as witnessed by comparison between the complete short-range expansion of the OBDM (Eq.~(4) of the main text) and the corresponding PIMC results.

The non-analytic cubic term, which is more important in the short-range behavior of the OBDM at distances approaching $x_{\rm max}$, is thermally faded at temperatures above the hole-anomaly threshold $T > T_A$ in any interacting regime, as reported by the comparison between the colored solid and black dashed lines.~This fading mechanism is a smooth crossover and not an abrupt change occurring exactly at the hole-anomaly temperature $T_A$.

\section{Momentum Distribution}
\label{Sec:n(k)}

\begin{figure}[h!]
\includegraphics[width=0.76\textwidth]{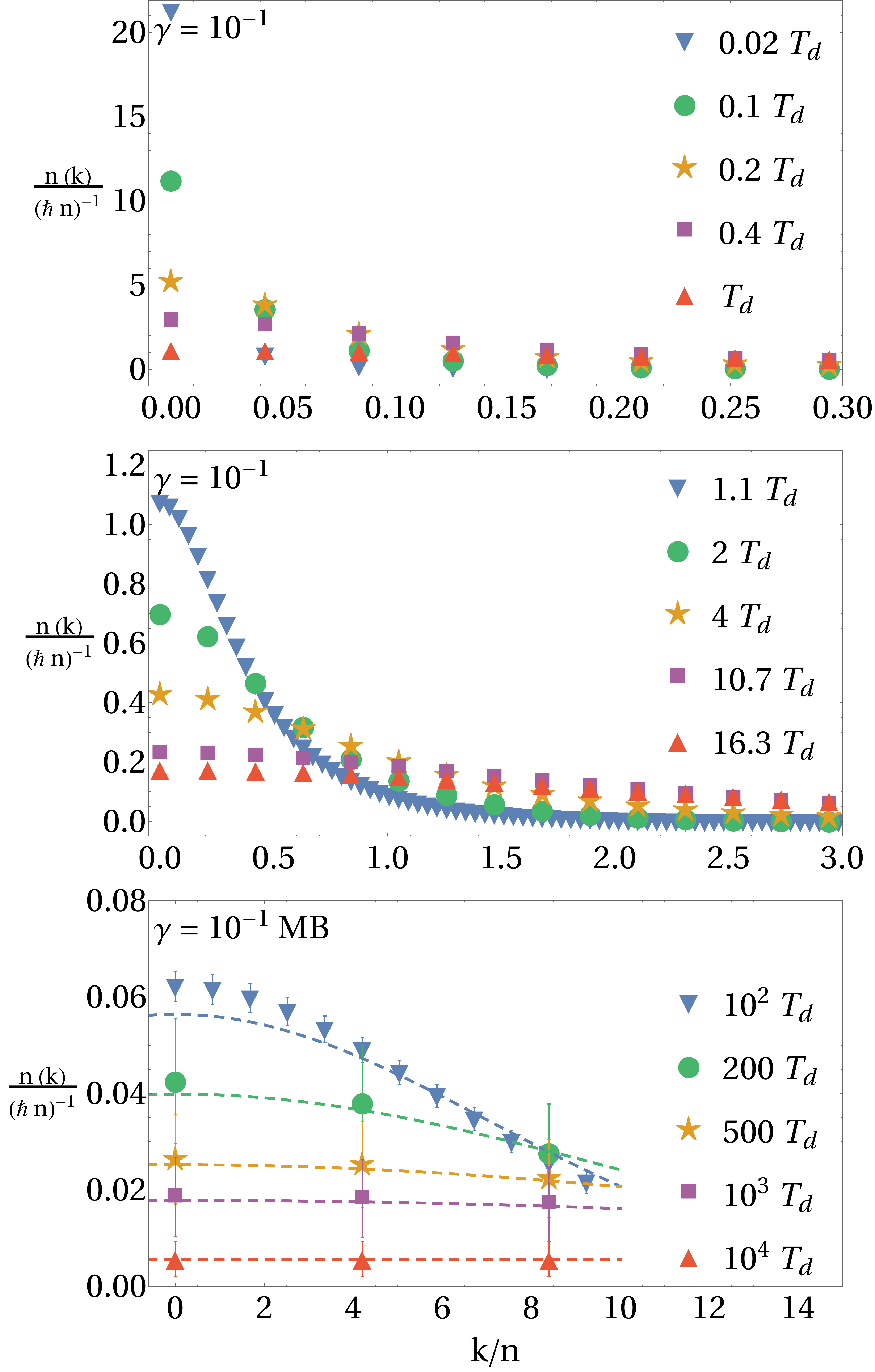}
\caption{Momentum distribution $n(k)/(\hbar n)^{-1}$ as a function of the momentum $k/n$ ($n$ is the density).~The weakly-interacting regime with $\gamma = 10^{-1}$ is considered.~Temperature is reported in units of the quantum degeneracy value $T_d = \hbar^2 n^2/\left(2 m k_B\right)$.~Symbols denote exact Path Integral Monte Carlo results and their sizes are larger than the statistical error bars in some cases.~Dashed lines represent the Maxwell-Boltzmann limit $n\left(k\right)_{\rm MB}/(\hbar n)^{-1} = \sigma/\left( \hbar \sqrt{2 \pi} \right) e^{- \sigma^2 k^2/2}/(\hbar n)^{-1}$ of the classical gas at high temperatures and are reported in increasing order of the temperature from small (top) to large (bottom) values in the lowest panel. 
}
\label{fig:nk01}
\end{figure}

\begin{figure}[h!]
\includegraphics[width=0.76\textwidth]{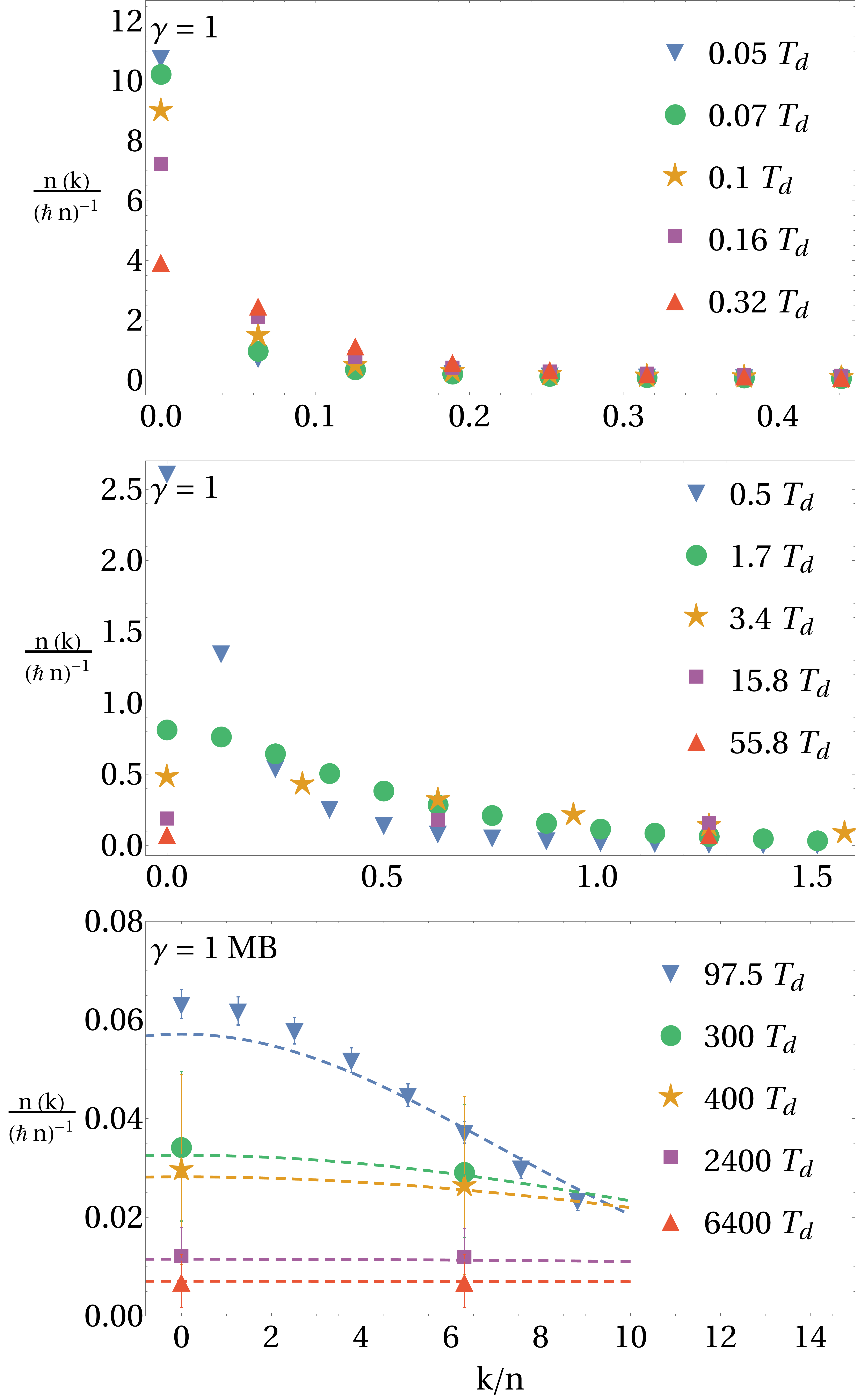}
\caption{Momentum distribution for the intermediate interaction strength $\gamma = 1$, similarly as in Fig.~\ref{fig:nk01}.~Symbols represent PIMC results.~Dashed lines denote the limit $n\left(k\right)_{\rm MB}/(\hbar n)^{-1}$ of the classical gas at very high temperatures and are reported in order of increasing temperature from small (top) to large (bottom) values in the lowest panel.  
}
\label{fig:nk1}
\end{figure}

\begin{figure}[h!]
\includegraphics[width=0.76\textwidth]{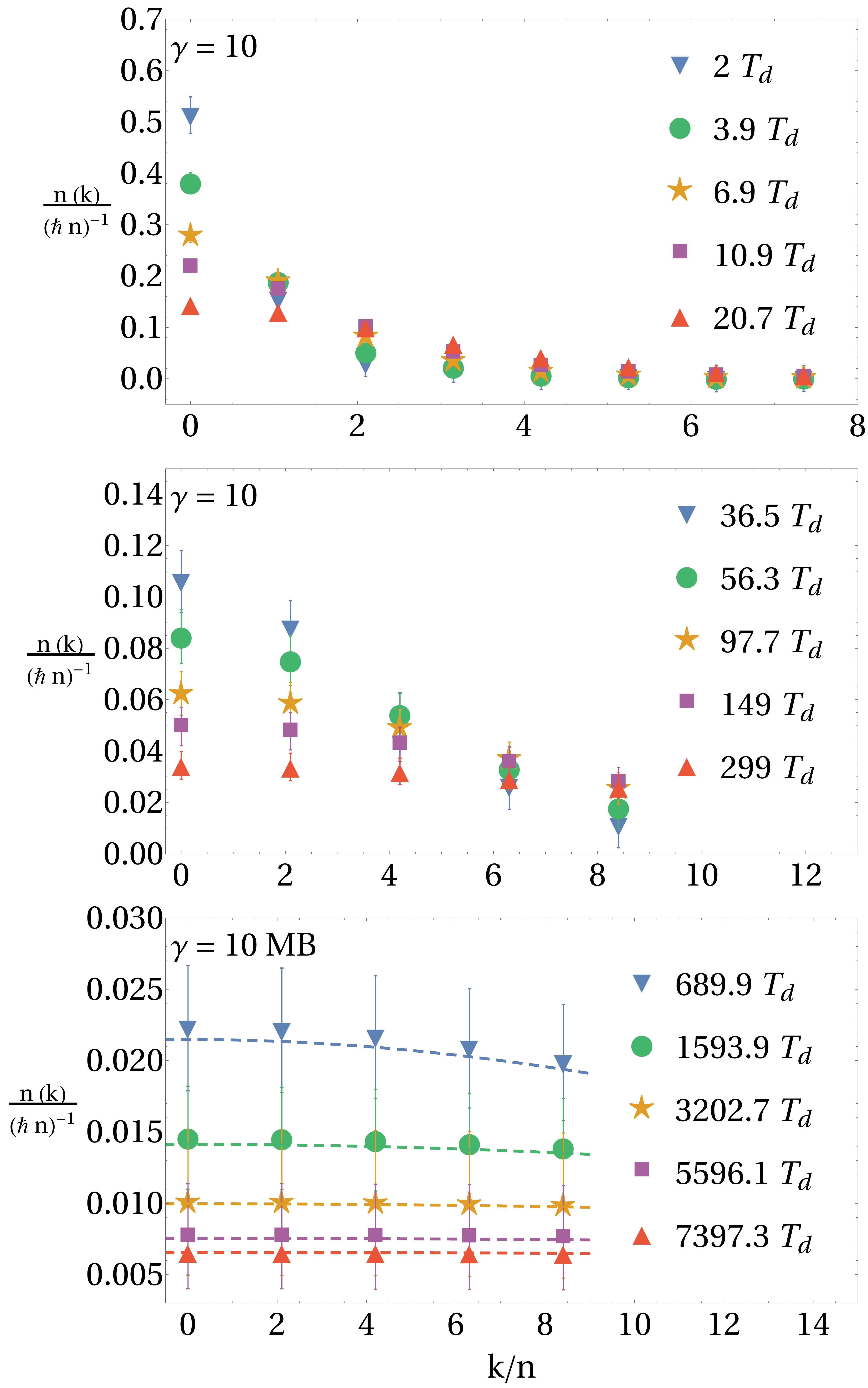}
\caption{Momentum distribution for strong interactions, $\gamma = 10$, similarly as in Fig.~\ref{fig:nk01}.~Symbols correspond to PIMC results.~Dashed lines represent the limit $n\left(k\right)_{\rm MB}/(\hbar n)^{-1}$ of the classical gas at high temperatures and are reported from low (top) to high (bottom) temperatures in the lowest panel. 
}
\label{fig:nk10}
\end{figure}

In Figs.~\ref{fig:nk01}-\ref{fig:nk10}, we present the complete $k$-dependence for the momentum distribution $n\left(k\right)$, by exploring all possible regimes in the interaction-temperature crossover.~For any value of the interaction strength $\gamma$, dashed lines in the lowest panels denote the analytical limit $n\left(k\right)_{\rm MB} = \sigma/\left( \hbar \sqrt{2 \pi} \right) e^{- \sigma^2 k^2/2}$, describing the classical gas at high temperatures, and which exhibits an excellent agreement with PIMC findings.

In Fig.~\ref{fig:nk01}, we present the case of weak interactions, $\gamma = 10^{-1}$, around the hole-anomaly (upper panel), intermediate temperatures (middle) and Maxwell-Boltzmann classical (lower) regime.

In Fig.~\ref{fig:nk1}, we report the intermediate value for the interaction strength $\gamma = 1$.~Each panel corresponds to different thermal regimes: very low temperatures and small momenta (upper panel), anomaly (middle) and MB (lower) regime.   

Fig.~\ref{fig:nk10} shows $\gamma = 10$ for strong repulsions, in the anomaly (upper panel), virial hard-core (middle) and MB (lower) regime.

At zero temperature and weak interparticle repulsion $\gamma \ll 1$, the occupation of the zero-momentum state features a power-law divergence with the particle number $N$: $n(k = 0) \propto N^{1-1/(2K)}$ where $K\geq 1$ is the Luttinger parameter \cite{Cazalilla2004}.~This is a remnant of the Bose-Einstein condensation, which is reached only asymptotically for $\gamma\to 0$ ($K \to \infty$), where $n(k=0)\propto N$.~At finite temperature, the peak in $n(k = 0)$ is smeared and the divergence is removed.~The peak gets smaller and broader with the increase of temperature and interaction strength, see Figs.~\ref{fig:nk01}-\ref{fig:nk10}.~For $\gamma = 10^{-1}$ and low temperatures, the peak is high, consistently with the presence of a quasicondensate \cite{Esteve2006}, but it does not show a delta-function behavior typical of a true condensate in higher spatial dimensions.~The thermal disappearance of the quasicondensate has been found even in other correlation properties \cite{DeRosi2022}.

\begin{figure}[h!]
\includegraphics[width=\textwidth]{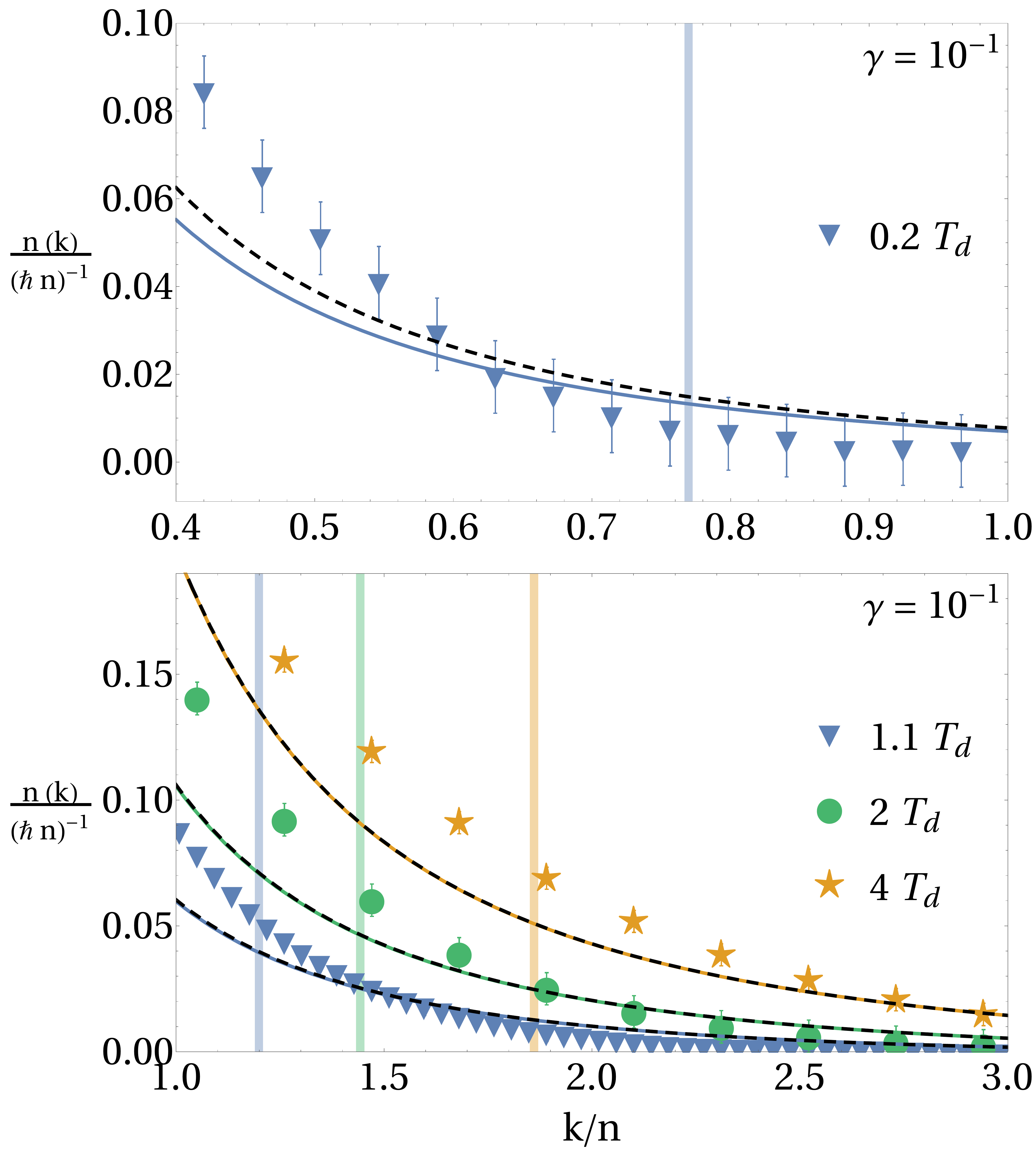}
\caption{Large-momentum behavior of the momentum distribution $n\left(k\right)$ in the weakly-interacting regime $\gamma = 10^{-1}$ and for temperatures below the hole-anomaly threshold $T < T_A$ (upper panel) and $T > T_A$ (lower) in units of the quantum degeneracy value $T_d = \hbar^2 n^2/\left(2 m k_B\right)$.~Symbols correspond to exact Path Integral Monte Carlo results.~Colored solid lines represent the complete tail at large momenta (Eq.~(9) of the main text) calculated with exact thermal Bethe-Ansatz method.~Dashed black lines correspond to the same, but with $b_3 = 0$.~All curves are reported in increasing order of the temperature from small (bottom) to large (top) values in the lower panel.~The minimum momentum for the validity of the distribution tail $k_{\rm min} = x_{\rm max}^{-1} = \left(\xi^{-1} + \sigma^{-1}\right)$ is represented with vertical lines with increasing order of temperature from small (left) to large (right) values. 
}
\label{fig:nkTail01}
\end{figure}

In Fig.~\ref{fig:nkTail01}, we report the distribution at large momenta $k$, for a weakly-repulsive gas $\gamma = 10^{-1}$ and temperatures below the anomaly threshold $T < T_A$ (upper panel) and at $T > T_A$ in the intermediate thermal regime (lower).~We compare the complete tail of the momentum distribution (Eq.~(9) of the main text), which is denoted by colored solid lines, with the same result with $b_3 = 0$ and corresponding to the black dashed lines.~Both curves are calculated with exact thermal Bethe-Ansatz method.~The minimum momentum $k_{\rm min} = x_{\rm max}^{-1} = \left(\xi^{-1} + \sigma^{-1}\right)$ for the validity of the complete tail corresponds to vertical lines.~Similar results in other interacting regimes are presented in the main text.

Consistently with the behavior of the short-distance OBDM in Sec.~\ref{Sec:OBDM}, $k_{\rm min}$ is valid for any interaction strength and temperature, as shown by the comparison between the complete tail (Eq.~(9) of the main text) and exact PIMC results in Fig.~\ref{fig:nkTail01} and similar plots in the main text.~We also notice that the new analytical approximation for the large-momentum tail of the distribution (Eq.~(9) of the main text) is valid in any interacting and thermal regime as it agrees with the corresponding PIMC data at momenta larger than the minimal value $|k| \gtrsim k_{\rm min}$.

\begin{figure}[h!]
\includegraphics[width=\textwidth]{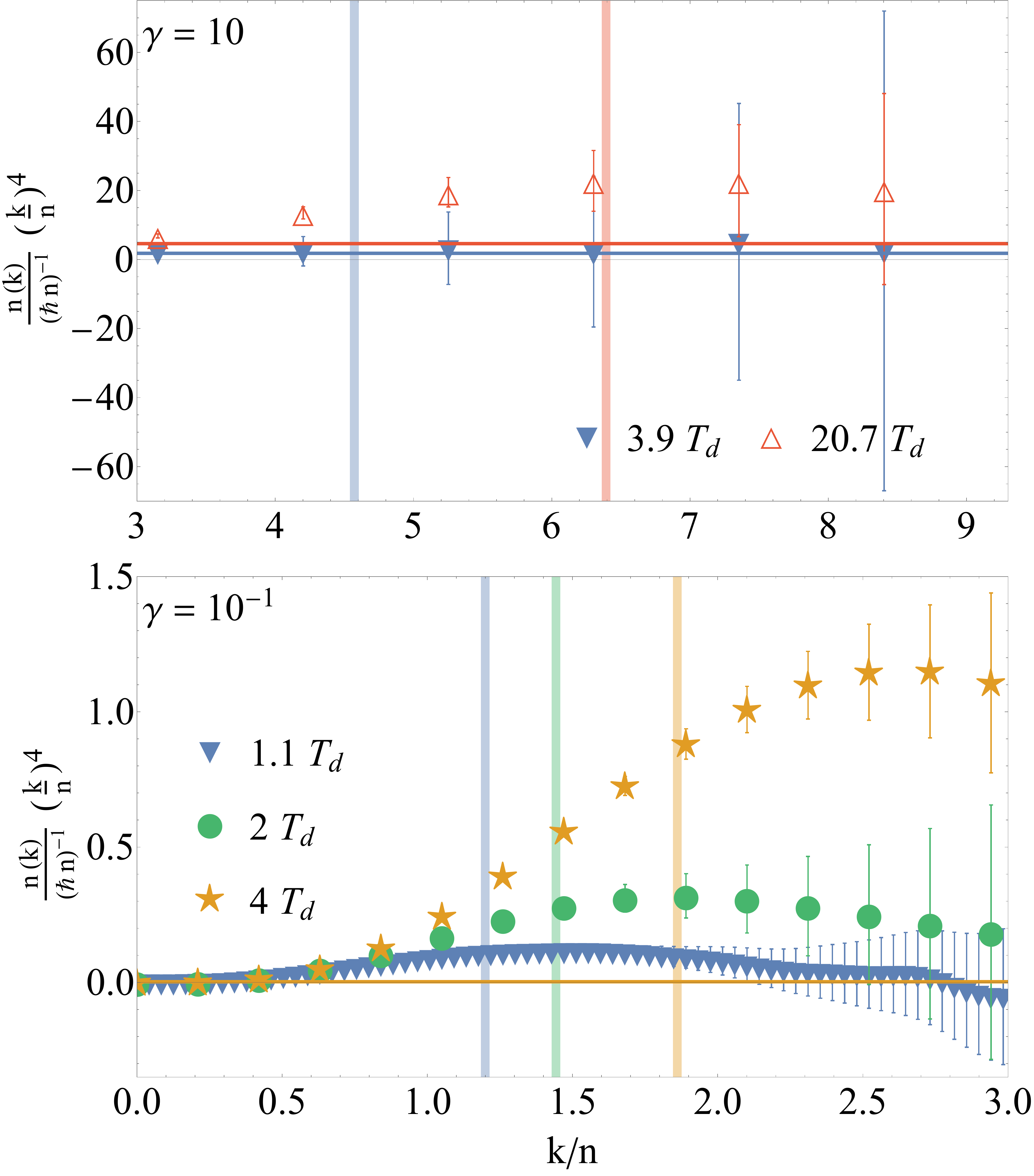}
\caption{Scaled momentum distribution $n\left(k\right) k^4$ as a function of the momentum $k$.~In the upper panel, we report $\gamma = 10$ for the strongly-interacting gas at temperatures below $T < T_A$ (solid symbols) and above $T > T_A$ (empty) the hole-anomaly threshold $T_A$.~In the lower panel, we show results for weak interactions $\gamma = 10^{-1}$ in the intermediate thermal regime with $T > T_A$.~Symbols correspond to exact Path Integral Monte Carlo results.~The minimum momentum for the distribution tail $k_{\rm min}$ is represented with vertical lines with increasing order of temperature from small (left) to large (right) values in each panel.~Horizontal lines report the coefficient of the Tan relation $6 n^3 b_3/\left(\pi \hbar \right)$ (see Eq.~(9) of the main text) calculated with exact thermal Bethe-Ansatz technique and they overlap at different temperatures in the lower panel. 
}
\label{fig:nk4k_SupMat}
\end{figure}

In Fig.~\ref{fig:nk4k_SupMat}, we present the scaled momentum distribution $n\left(k\right) k^4$ to make even more visible the thermal fading of the Tan relation $\mathcal{C}/k^4$ (where $\mathcal{C}$ is the Tan contact) in the large-$k$ tail at high temperatures.~In the upper panel, we consider the strongly-interacting regime $\gamma = 10$ and results for $T < T_A$ ($T > T_A$) reported with solid (empty) symbols.~In the lower panel, we show instead the weakly-repulsive system with $\gamma = 10^{-1}$ and in the intermediate thermal regime $T  > T_A$.~$k_{\rm min}$ is denoted by vertical lines.~Horizontal lines correspond to the coefficient of the Tan-relation tail $6 n^3 b_3/\left(\pi \hbar \right) \sim \mathcal{C}$ (see Eq.~(9) of the main text) which is computed with thermal Bethe-Ansatz method.~Additional findings for other values of the interaction strength $\gamma$ and temperatures are reported in the main text.

The thermal fading of the Tan relation in the tail of the momentum distribution takes place for any finite contact-interaction strength and temperatures above the hole-anomaly threshold $T > T_A$ including the typical values of the classical gas regime.~The fading crossover is more evident by raising the temperature and at momenta close to $k_{\rm min}$, where the contribution of the Tan relation is important in the analytical approximation for the complete tail (Eq.~(9) of the main text).

\bibliography{Bibliography}